\documentclass[3p,authoryear,11pt]{elsarticle}

\usepackage{amssymb}
\usepackage{amsmath}
\usepackage{stmaryrd}
\usepackage{url}
\usepackage{comment}
\usepackage{array}

\newcommand{\vect}[1]{\boldsymbol{#1}}

\newcommand{\eps}{\varepsilon}

\newcommand{\dif}{\, d}

\newcommand{\pd}{\partial}

\newcommand{\brac}[1]{\left( {#1} \right)}
\newcommand{\bracc}[1]{\left\{ {#1} \right\}}
\newcommand{\bracs}[1]{\left[ {#1} \right]}

\newcommand{\jump}[1]{\llbracket {#1} \rrbracket}
\newcommand{\avg}[1]{\langle {#1} \rangle}

\let\oldmarginpar\marginpar
\renewcommand\marginpar[1]{\-\oldmarginpar[\raggedright\scriptsize #1]{\raggedright\scriptsize #1}}
\begin{document}
\begin{frontmatter}
\title{Phase field model for coupled displacive and diffusive
microstructural processes under thermal loading}

\author[mm]{Mirko Maraldi}
\ead{mirko.maraldi@unibo.it}
\address[mm]{DIEM, University of Bologna,
              V.le Risorgimento, 2, 40136 Bologna, Italy}

\author[gnw]{Garth N. Wells\corref{cor1}}
\ead{gnw20@cam.ac.uk}
\address[gnw]{Department of Engineering, University of Cambridge,
              Trumpington Street, Cambridge CB2 1PZ, United Kingdom}

\author[lm]{Luisa Molari}
\ead{luisa.molari@unibo.it}
\address[lm]{DICAM, University of Bologna,
              V.le Risorgimento, 2, 40136 Bologna, Italy}

\cortext[cor1]{Corresponding author}

\begin{abstract}
A non-isothermal phase field model that captures both displacive and
diffusive phase transformations in a unified framework is presented.
The model is developed in a formal thermodynamic setting, which provides
guidance on admissible constitutive relationships and on the coupling of
the numerous physical processes that are active.  Phase changes are driven
by temperature-dependent free-energy functions that become non-convex
below a transition temperature.  Higher-order spatial gradients are
present in the model to account for phase boundary energy, and these
terms necessitate the introduction of non-standard terms in the energy
balance equation in order to satisfy the classical entropy inequality
point-wise.  To solve the resulting balance equations, a Galerkin finite
element scheme is elaborated. To deal rigorously with the presence of
high-order spatial derivatives associated with surface energies at phase
boundaries in both the momentum and mass balance equations, some novel
numerical approaches are used.  Numerical examples are presented that
consider boundary cooling of a domain at different rates, and these
results demonstrate that the model can qualitatively reproduce the
evolution of microstructural features that are observed in some alloys,
especially steels.  The proposed model opens a number of interesting
possibilities for simulating and controlling microstructure pattern
development under combinations of thermal and mechanical loading.
\end{abstract}
\end{frontmatter}
\section{Introduction}
\label{sec:introduction}

Displacive and diffusive phase transformations in solids, driven by
temperature changes or mechanical loading, occur in many industrial
and natural processes.  In particular, careful temperature control is
used extensively to architect microstructural features of alloys and
thereby tailor their mechanical properties. We present in this work
a non-isothermal phase field model for simulating the development of
microstructure due to both displacive and diffusive processes under
thermal and mechanical loading. Continuum-level order parameters are
defined that indicate the presence of various phases, and coupled
evolution equations for these parameters are developed. The model is
framed in a thermodynamic setting which provides guidance on the coupling
of various processes.

Displacive phase transitions have been extensively studied over a long
period.  \citet{wechsler:1953} and \citet{bowles:1954} proposed theories
of cubic-to-tetragonal martensitic phase transitions in which the key
feature is the description of the `Bain distortion' \citep{bhadeshia:1987,
wayman:1990}. The formation of martensite twins has also been studied
in the context of energy minimisers \citep{ball:1987, kohn:1991,
bhattacharya:1991}.  \citet{falk:1980} proposed a model based on a
single order parameter, with a free-energy that is a non-convex sextic
polynomial in the order parameter, although the model did not include
evolution equations for the order parameter or surface (phase boundary)
energy contributions to the free-energy. It is not uncommon that such
singular models are studied, see for example \citet{ball:2004}. However,
models that do not account for surface energy cannot predict the detail of
the microstructure, which is determined by the relative balance between
bulk and interface energy, and such models will generally lead to an
ill-posed problem when inserted into a differential balance equation.
The above references do not consider evolution equations, in which case
kinetic aspects of a transition are ignored.  \citet{barsch:1984}
and \citet{jacobs:1985} proposed models featuring both bulk and
interfacial contributions to the free-energy, together with balance
of linear momentum, thereby addressing the evolution of transitions.
Our treatment of the displacive transformations is in the same spirit
as \citet{barsch:1984} and \citet{jacobs:1985}.  There exist other
phase field approaches to martensitic transformations that include
bulk and surface energy, but that do not invoke balance of momentum
\citep{wang:1997,shenoy:1999,onuki:1999}.  These models involve the
solution of a diffusion-type problem.

For diffusive transformations, the phase field approach has been
used extensively to model microstructure evolution.  The best known
model for conserved order parameters is the Cahn-Hilliard equation
\citep{cahn:1958}, in which surface energy is introduced via spatial
gradients of the order parameter in the free-energy. Of relevance to the
problems that we will consider, the Cahn-Hilliard equations was extended
to include non-isothermal effects in \citet{alt:1992}.

In the heat treatment of steel alloys, both displacive and diffusive
transformation are important.  Displacive transformations (martensite
formation) occur rapidly compared to diffusive transformations
(pearlite formation), with the speed of latter being limited by
the rate at which carbon atoms can diffuse.  There have been many
studies into models for displacive or diffusive phase transitions,
but fewer into models that can capture both displacive and diffusive
transformations, and interactions between the resulting phases.
\citet{wang:1993} presented an isothermal model Ginzburg-Landau type
model for simulating both displacive and diffusive transformations via
coupled Cahn-Allen and Cahn-Hilliard type equations.  Consistent with
the Ginzburg-Landau framework, the model of \citet{wang:1993} does
not invoke balance of momentum.  \citet{bouville:2006} presented an
isothermal model for coupled displacive/diffusive processes (see also
\citet{bouville:2007}), in which the treatment of displacive processes
resembles that of \citet{barsch:1984}, and diffusive processes are
modelled by a Cahn-Hilliard type equation. Interactions between processes
was accounted for via coupling terms in the free-energy, and the impact
of various coupling terms was studied numerically.

While the heat treatment of steel is perhaps the most technologically
relevant, diffusive and displacive transformations driven by mechanical
and/or temperature loading are relevant for a variety of materials. For
example, experimental observations of martensitic twinning in perovskites
are presented by~\citet{harrison:2004}.  Our intention in this work is to
present a generic framework and to demonstrate that the development of key
microstructural features can be modelled. Because of the degree to which
heat treatment is used for steels, we will at times use steel-specific
terminology and refer to a high temperature stable phase as austenite
and a diffusive phase as `pearlite', despite the generic nature of
the formulation.

A focus of this work is the formulation of a non-isothermal model in a
thermodynamic setting.  Free-energy expressions that involve non-standard
higher-order gradient terms to model surface energies are postulated,
and a model that satisfies the first and second laws of thermodynamics
is formulated.  Dealing with the higher-order gradients in a classical
thermodynamic setting is not trivial, so in formulating our model we adopt
an approach with parallels to that advocated by \citet{gurtin:1996},
and we show that the concept of `work done' by higher-order (nonlocal)
terms on the boundary of a domain is necessary to formulate a model that
satisfies the entropy inequality point-wise.  Many of the aforementioned
works on phase field models invoke thermodynamic arguments, but usually
premised on the \emph{a priori} postulation of a chemical potential and
proportionality between a mass flux and the chemical potential. Following
\emph{ad-hoc} arguments \citet{maraldi:2010} include the heat
equation in their phase field model, following the usual methodology of
assuming a structure for the stress tensors and the chemical potential.
A distinguishing feature of our formulation is that it follows from an
energy balance that is posed in terms of the work done on the boundary
of a domain, in the spirit of \citet{gurtin:1996}.  The equations that
result from our derivation are not trivial to solve. In particular, the
incorporation of surface energies via higher-order spatial gradients
leads to coupled fourth-order hyperbolic and parabolic equations.
We therefore pay attention to the development of a Galerkin finite
element formulation of the equations, which we then use to compute a
number of example problems.  The presented formulation is novel in the
coupling of displacive, diffusive and thermal processes. Moreover it is
distinguished from other works by the formal thermodynamic framework in
which the model is developed, and the sophistication of the numerical
method formulated for solving the resulting equations.

The rest of this work is structured as follows. The scale of the
problem that we consider is discussed and order parameters that are
relevant at the considered scale are defined.  This is followed by
the formulation of thermodynamic balance laws and formalisation of
the restrictions imposed by the entropy inequality. Balance laws and
constitutive models are then synthesised to yield governing equations,
which is followed by the development of a suitable Galerkin finite
element method.  A number of simulations that qualitatively illustrate
features of non-isothermal coupled phase transformations are presented.
The computer code used to produce all numerical examples is freely
available and distributed under a GNU public license. It can be found in
the supporting material~\citep{supporting_material:2010}.  Following the
numerical examples, some conclusions are drawn.

\section{Order parameters and scale of the problem}
\label{sec:order_parameters}

We define now order parameters whose values will be used to identify
phases. The choice of order parameters depends on the scale of
observation. We consider the level of a single grain, at which scale
the lattice orientation can be inferred, but at which the continuum
hypothesis remains valid. The domain of interest will be denoted by
$\Omega \subset \mathbb{R}^{d}$, where $d$ is the spatial dimension.

Diffusive-type transformations will be characterised by a conserved
scalar order parameter~$c$.  The parameter $c$ typically represents the
relative concentration of an alloying element, such as carbon, and can act
as an indicator of phase. In steels, for example, if $c$ represents the
deviation in the carbon concentration away the equilibrium concentration
in austenite, then the presence of pearlite is indicated by regions in
which $c$ alternates spatially about zero.  The carbon-rich cementite
phase can be identified by positive $c$, and the ferrite phase can be
identified by negative~$c$.

To characterise displacive transformations, deformation-related order
parameters are considered.  In particular, we will use scalar order
parameters $e_{i}$ that are functions of the linearised strain tensor
$\vect{\eps}= \brac{\nabla \vect{u} + \brac{\nabla \vect{u}}^{T}}/2$,
where $\vect{u}$ is the displacement vector.  The formation of phases
driven by displacive effects at the scale of observation that we have
chosen is dependent on the lattice orientation.  We restrict ourselves to
two spatial dimensions ($d = 2$) and an initially square lattice aligned
with the Cartesian $x_{1}$ and $x_{2}$ axes, in which case we consider
a volumetric order parameter~$e_{1}$,
\begin{equation}
  e_{1} = \eps_{11} + \eps_{22},
\end{equation}
an order parameter $e_{2}$,
\begin{equation}
  e_{2} = \eps_{11} - \eps_{22},
\end{equation}
and the shear strain $e_{3}$,
\begin{equation}
  e_{3} = \eps_{12}.
\end{equation}
The order parameter $e_{2}$ acts as an indicator of a martensitic
phase, with a significant variation away from zero indicating a
square-to-rectangle type transition \citep{jacobs:1985}. Martensite
twins will be evident when $e_{2}$ alternates spatially about zero.
\section{Thermodynamic balance laws and restrictions}
\label{sec:thermodynamics}

To formulate thermodynamic restrictions on constitutive equations, it
is sufficient at this stage to postulate the existence of a suitably
smooth Helmholtz free-energy density functional $f$ of the form
\begin{equation}
  f = f\brac{\vect{\eps}, \nabla \vect{\eps}, c, \nabla c, T},
\label{eqn:f_functional_form}
\end{equation}
where $T$ is the temperature.  A dependency of the free-energy on the
gradient terms $\nabla \vect{\eps}$ and $\nabla c$ is included with a
view to the inclusion of surface energies in the context of a diffuse
interface model.  A precise form of the free-energy as a function of
the order parameters introduced in Section~\ref{sec:order_parameters}
will be presented in Section~\ref{sec:free_energy}.

Special care is required in constructing the energy balance in order
to properly account for mass diffusion and the presence of higher-order
spatial gradients in the free-energy.  We adopt an approach to the energy
balance that shares features with the approach of \citet{gurtin:1996}
for the Cahn-Hilliard equation, in which non-standard force-like terms
associated with higher order terms are identified and accounted for in
the energy balance equation.  Unlike in the work of \citet{gurtin:1996},
we do not specify \emph{a priori} any new balance laws for these extra
terms, but we will show that balance laws for these terms are implied
by insisting upon satisfaction of the classical entropy inequality.

We will consider linearised kinematics throughout. The open domain $R
\subseteq \Omega$ is used to denote an arbitrary sub-region of~$\Omega$,
hence integral balance laws posed on $R$ can be localised. The outward
unit normal vector to $R$ (on $\pd R$) is denoted by~$\vect{n}$.
Since $c$ is assumed to be a conserved order parameter it must satisfy
\begin{equation}
  \frac{d}{dt}\int_{R} c \dif x = - \int_{R} \nabla \cdot \vect{j}_{c} \dif x,
\label{eqn:mass_balance_integral}
\end{equation}
where $\vect{j}_{c}$ is the mass flux of~$c$.  The classical linear
momentum balance equation reads:
\begin{equation}
  \frac{d}{dt} \int_{R} \rho \vect{v} \dif x
    =  \int_{R} \nabla \cdot \vect{\sigma} \dif x
    +  \int_{R}  \vect{b} \dif x,
\label{eqn:momentum_balance_integral}
\end{equation}
where $\rho$ is the mass density, assumed to be constant,
$\vect{v}$ is the velocity, $\vect{\sigma} = \vect{\sigma}^{T}$
is the stress and $\vect{b}$ is a body force. Localising
equation~\eqref{eqn:momentum_balance_integral},
\begin{equation}
   \rho \frac{d \vect{v}}{dt}
    =  \nabla \cdot \vect{\sigma} +  \vect{b}.
\label{eqn:momentum_balance_local}
\end{equation}
Multiplying the balance of linear momentum
equation~\eqref{eqn:momentum_balance_local} by $\vect{v}$, and then
integrating over $R$ and applying integration by parts leads to the
mechanical energy balance:
\begin{equation}
  \frac{1}{2} \frac{d}{dt} \int_{R} \rho \| \vect{v} \|^{2} \dif x =
      - \int_{R}  \vect{\sigma} : \nabla \vect{v}\dif x
      + \int_{\pd R}  \vect{t} \cdot \vect{v}  \dif s
      + \int_{R}  \vect{b} \cdot \vect{v} \dif x,
\label{eqn:mech_power_balance}
\end{equation}
where $\vect{t} = \vect{\sigma}\vect{n}$ is the traction.

We now consider an energy balance equation of the form
\begin{multline}
	 \frac{d}{dt} \int_{R} \frac{1}{2} \rho \| \vect{v} \|^{2} + u \dif x
   =
  - \int_{\pd R} \vect{q} \cdot \vect{n} \dif s
  + \int_{\pd R} \vect{t} \cdot \vect{v} \dif s
  + \int_{R} \vect{b} \cdot \vect{v} \dif  x
\\
  + \int_{\pd R} \vect{\Sigma} \vect{n} : \Dot{\vect{\eps}}  \dif s
  + \int_{\pd R} \vect{\xi} \cdot \vect{n} \Dot{c}  \dif s
  - \int_{\pd R} \mu \vect{j}_{c} \cdot \vect{n} \dif s,
\label{eqn:energy_balance}
\end{multline}
where $u$ is the specific internal energy density (per unit volume),
$\vect{q}$ is the heat flux, the third-order tensor $\vect{\Sigma}$ and
the vector $\vect{\xi}$ are stress-like terms, and $\mu \vect{j}_{c}$
is related to the energy transported into the domain by the mass flux
of~$c$.  The terms $\vect{\Sigma}$ and $\vect{\xi}$ are not standard,
and their presence is a consequence of the nonlocality implied by the
dependency of the free-energy on $\nabla \vect{\eps}$ and~$\nabla c$. The
`fluxes' $\vect{\Sigma}\vect{n} : \Dot{\vect{\eps}}$ and $\vect{\xi}
\cdot \vect{n} \Dot{c}$, evaluated on $\pd R$, can be interpreted as
the power expended across the boundary of $R$ by a material particle
just outside~$R$.  This concept is used by \citet{gurtin:1996} for the
Cahn-Hilliard equation, and less explicitly by \citet{polizzotto:2003}
for gradient elasticity problems. \citet{polizzotto:2003} introduces
the concept of a `nonlocal' residual in the energy balance equation to
account the non-standard terms that arise in gradient elasticity. The
precise role of $\vect{\Sigma}$ and $\vect{\xi}$ will become more
evident when considering admissible constitutive equations.  Inserting
the mechanical energy balance~\eqref{eqn:mech_power_balance} into the
energy balance~\eqref{eqn:energy_balance} and applying the divergence
the theorem leads to a local internal energy balance equation:
\begin{equation}
	 \Dot{u}
   =
   - \nabla \cdot\vect{q}
   + \vect{\sigma} : \nabla \vect{v}
   + (\nabla \cdot \vect{\Sigma}) : \nabla \vect{v}
   + \vect{\Sigma} : \nabla\Dot{\vect{\eps}}
\\
   + (\nabla \cdot \vect{\xi}) \Dot{c}
   + \vect{\xi} \cdot \nabla \Dot{c}
   - \mu \nabla \cdot \vect{j}_{c}
   - \nabla \mu \cdot \vect{j}_{c}.
\label{eqn:internal_energy_balance}
\end{equation}

We will insist upon the satisfaction of a local entropy inequality of
the form
\begin{equation}
  \dot{s}
 \ge
- \nabla \cdot \brac{\frac{\vect{q}}{T}},
\label{eqn:entropy_inequality}
\end{equation}
where $s$ is the entropy density per unit volume.  Using the
definition of the Helmholtz free-energy ($f = u - Ts$) in the entropy
inequality~\eqref{eqn:entropy_inequality} leads to
\begin{equation}
  \Dot{u} - \Dot{f} - \Dot{T} s
        + T \nabla \cdot\brac{\frac{\vect{q}}{T}}
  \geq  0.
\end{equation}
Inserting the internal energy balance~\eqref{eqn:internal_energy_balance}
into the above inequality yields
\begin{multline}
  -\nabla \cdot \vect{q}
     + \vect{\sigma} : \nabla\vect{v}
     + \nabla \cdot \vect{\Sigma} : \Dot{\vect{\eps}} + \vect{\xi} : \nabla \Dot{\vect{\eps}}
     + \nabla \cdot \vect{\xi} \Dot{c} + \vect{\xi} \cdot \nabla \Dot{c}
\\
     + \mu \dot{c}
     - \nabla \mu \cdot \vect{j}_{c}
     - \Dot{f} - \Dot{T} s
     + T \nabla \cdot\brac{\frac{\vect{q}}{T} }
  \geq  0,
\end{multline}
which can be manipulated into the form
\begin{equation}
     \vect{\sigma} : \nabla \vect{v}
     + \nabla \cdot \vect{\Sigma} : \Dot{\vect{\eps}} + \vect{\xi} : \nabla \Dot{\vect{\eps}}
    + \nabla \cdot \vect{\xi} \Dot{c} + \vect{\xi} \cdot \nabla \Dot{c}
    + \mu \dot{c}
\\
     - \nabla \mu \cdot \vect{j}_{c}
     - \Dot{f} - \Dot{T} s
     - \nabla T \cdot \brac{\frac{\vect{q}}{T} }
  \geq  0.
\label{eqn:entropy_inequality_constit}
\end{equation}
We will insist upon satisfaction of this inequality in our model.

\section{Admissible constitutive equations}
\label{sec:admissible_constitutive_equations}
By insisting upon satisfaction of the entropy
inequality~\eqref{eqn:entropy_inequality_constit}, the precise form of
some constitutive models will follow directly from the definition of the
free-energy, while for the others it will simply imply a restriction.
To start, for a Helmholtz free-energy that has the functional form of
equation~\eqref{eqn:f_functional_form}, its time derivative reads:
\begin{equation}
  \Dot{f}
=  \frac{\pd f}{\pd T} \Dot{T}
  + \frac{\pd f}{\pd \vect{\eps}} : \Dot{\vect{\eps}}
  + \frac{\pd f}{\pd \nabla \vect{\eps}} \vdots \nabla \Dot{\vect{\eps}}
  + \frac{\pd f}{\pd c} \Dot{c}
  + \frac{\pd f}{\pd \nabla c} \cdot \nabla \Dot{c}.
\label{eqn:free_energy_rate}
\end{equation}
Inserting the above expansion of $\Dot{f}$ into the entropy
inequality~\eqref{eqn:entropy_inequality_constit},
\begin{multline}
  \brac{\vect{\sigma}_{e} + \nabla \cdot \vect{\Sigma}
        - \frac{\pd f}{\pd \vect{\eps}}} : \Dot{\vect{\eps}}
    + \vect{\sigma}_{v} : \Dot{\vect{\eps}}
+\brac{\vect{\Sigma} - \frac{\pd f}{\pd \nabla \vect{\eps}}} : \nabla \Dot{\vect{\eps}}
     - \brac{s + \frac{\pd f}{\pd T}}  \Dot{T}
\\
     + \brac{\nabla \cdot \vect{\xi} + \mu - \frac{\pd f}{\pd c} } \Dot{c}
     + \brac{\vect{\xi} - \frac{\pd f}{\pd \nabla c}} \cdot \nabla \Dot{c}
     - \nabla\mu \cdot \vect{j}_{c}
        - \nabla T  \cdot \brac{\frac{\vect{q}}{T}}
  \geq  0,
\label{eqn:entropy_inequality_f}
\end{multline}
where is has been assumed that the stress tensor can can be
decomposed additively into inviscid ($\vect{\sigma}_{e}$) and viscous
($\vect{\sigma}_{v}$) parts,\footnote{This permits the incorporation
of a Kelvin-Voigt type model. Other models that involve springs and
dashpots in series can be formulated via the introduction of strain-like
internal variables.}
\begin{equation}
  \vect{\sigma} = \vect{\sigma}_{e} + \vect{\sigma}_{v}.
\end{equation}
From the arbitrariness of $\vect{v}$, $c$ and $T$, and the insistence
upon satisfaction of equation~\eqref{eqn:entropy_inequality_f},
we can infer various admissible constitutive relationships.
Equation~\eqref{eqn:entropy_inequality_f} implies for the `higher-order'
stress $\vect{\Sigma}$ that
\begin{equation}
  \vect{\Sigma} =  \frac{\pd f}{\pd \nabla \vect{\eps}},
\label{eqn:Sigma_definition}
\end{equation}
for the inviscid component of the stress that
\begin{equation}
  \vect{\sigma}_{e}
    = \frac{\pd f}{\pd \vect{\eps}} - \nabla \cdot \vect{\Sigma}
    = \frac{\pd f}{\pd \vect{\eps}} - \nabla \cdot \frac{\pd f}{\pd \nabla \vect{\eps}},
\label{eqn:stress_definition}
\end{equation}
and that a constitutive model for the viscous stress must satisfy
\begin{equation}
    \vect{\sigma}_{v} : \nabla \vect{v} \ge 0.
\label{eqn:viscous_stress_restriction}
\end{equation}
Satisfaction of equation~\eqref{eqn:entropy_inequality_f} also
requires that the `chemical higher-order stress' be given by
\begin{equation}
  \vect{\xi} =  \frac{\pd f}{\pd \nabla c},
\end{equation}
that the `chemical potential' is given by
\begin{equation}
  \mu = \frac{\pd f}{\pd c} - \nabla\cdot\vect{\xi}
      = \frac{\pd f}{\pd c} - \nabla \cdot \frac{\pd f}{\pd \nabla c},
\label{eqn:mu_definition}
\end{equation}
and that a constitutive model for the mass flux vector must satisfy
\begin{equation}
  \nabla \mu \cdot \vect{j}_{c} \le 0.
\label{eqn:mass_flux_restriction}
\end{equation}
The entropy density is given by
\begin{equation}
  s = - \dfrac{\pd f}{\pd T}.
\label{eqn:entropy_definition}
\end{equation}
For the heat flux, the entropy inequality requires that
\begin{equation}
   \nabla T \cdot \brac{\frac{\vect{q}}{T} } \le 0.
\label{eqn:heat_flux_restriction}
\end{equation}

Without the inclusion of the non-standard terms $\vect{\Sigma}$ and
$\vect{\xi}$ in the energy balance, the classical entropy inequality
could not be satisfied point-wise, as shown in \citet{gurtin:1965} for
elasticity.  This is evident in equation~\eqref{eqn:entropy_inequality_f},
which in the absence of $\vect{\Sigma}$ and $\vect{\xi}$ could not be
guaranteed to hold since the terms $\pd f/ \pd \nabla\vect{\eps}$ and
$\pd f/ \pd \nabla c$ would be without a `partner' stress term.
\section{Specific form of the Helmholtz free-energy and constitutive equations}
\label{sec:free_energy}

Before presenting the boundary value problems that define the complete
model, it is useful to specify more precisely the functional form of
a Helmholtz free-energy density in terms of the order parameters that
is suitable for modelling phase transformations. It also useful to
define the adopted constitutive models that do not follow as a direct
consequence of the chosen free-energy expression.  While the governing
equations will be presented in a format that is largely independent
of the details of the free-energy function, some poignant features of
the governing equations only become apparent after the introduction
of particular constitutive equations, especially those related to the
surface energy. Where convenient, we will express constitutive models in
terms of derivatives of the free-energy.  This is because the presented
numerical simulations employ novel techniques for the automated generation
of computer code from a domain-specific language that can compute the
necessary derivatives automatically. The computer code therefore requires
expressions for the free-energy only.

\subsection{Helmholtz free-energy}
\label{sec:Helmholtz-free-energy}
We consider a Helmholtz free-energy density that can be additively
composed according to
\begin{equation}
f
	=   f_{\rm diff}\brac{T, c, \nabla c}
    + f_{\rm disp}\brac{T, e_{1}, e_{2}, e_{3}, \nabla e_{2}, c}
    + f_{\rm cpl}\brac{c, e_{2}}
    + f_{\rm therm}\brac{T},
\label{eqn:free-energy-decomposition}
\end{equation}
where $f_{\rm diff}$ is the free-energy associated with diffusive
processes, $f_{\rm disp}$ is the free-energy associated with displacive
transformations and mechanical deformation, $f_{\rm cpl}$ is the
free-energy associated with the interaction of phases and $f_{\rm therm}$ is
a part of the free-energy which is dependent on the temperature only. The
strain-related order parameters $e_{i}$ and the mass concentration order
parameter $c$ were defined in Section~\ref{sec:order_parameters}.
The functional forms of $f_{\rm diff}$, $f_{\rm disp}$ and $f_{\rm cpl}$
that we adopt come from \citet{bouville:2006}, with some minor modifications.

\subsubsection{Diffusive part of the free-energy}
\label{sec:diffusive_free_energy}
\begin{figure}
  \center\includegraphics[scale=0.6]{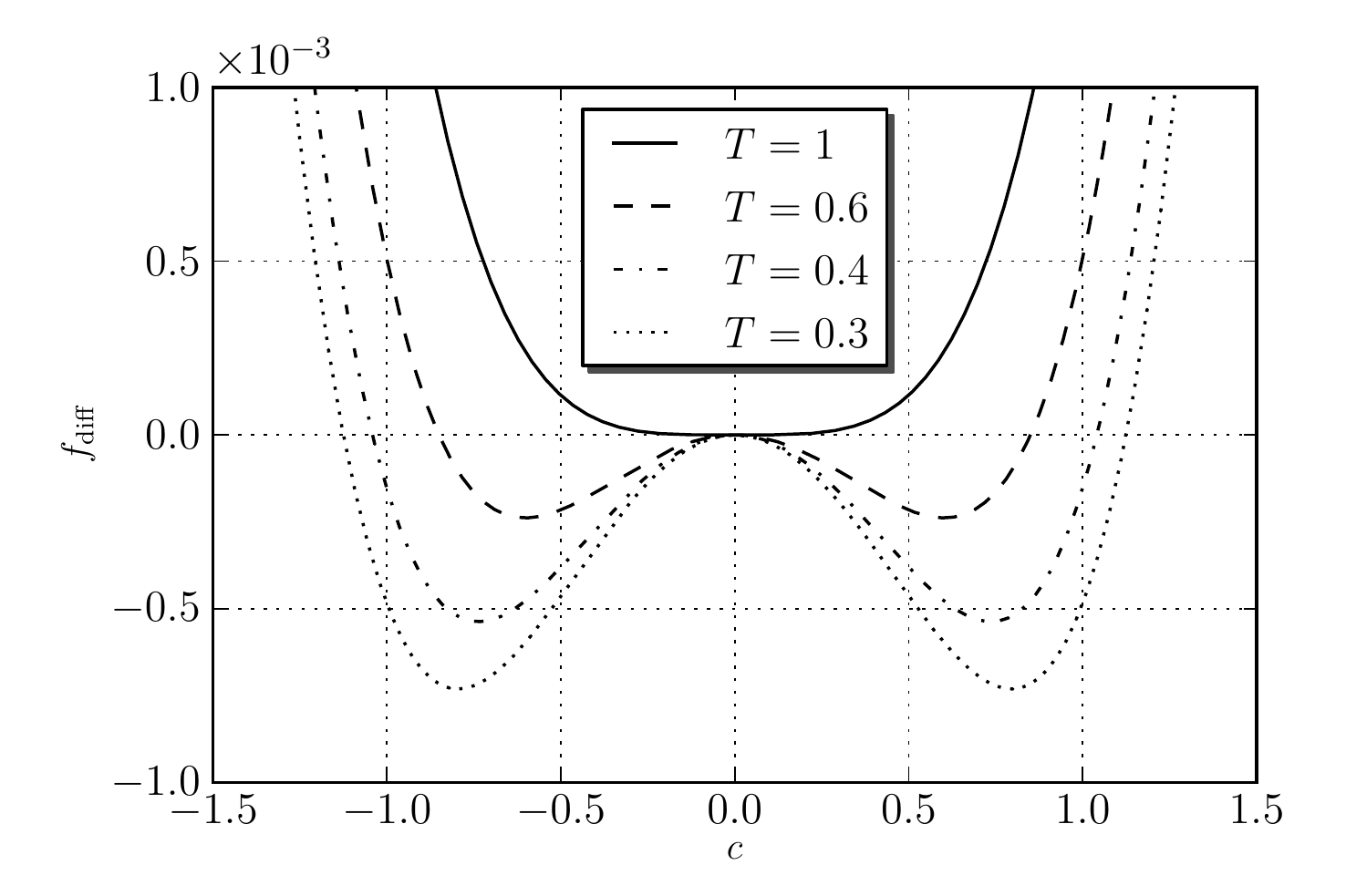}
\caption{Diffusive free-energy density as a function of $c$ at various
temperatures $T$ ($A_{4} = 7.31 \times 10^{-3}$, $A_{2} = 6.62 \times
10^{-3}$, $T_{P}=1$).}
\label{fig:diff_free_energy}
\end{figure}

We postulate a diffusive free-energy function of the form
\begin{equation}
  f_{\rm diff}
      =  \dfrac{A_{4}}{4} c^{4} + \dfrac{A_{2}}{2} \dfrac{T - T_{P}}{T_{P}} c^{2}
      +  \frac{\lambda_{c}}{2}  \| \nabla c \|^2,
\label{eqn:free_energy_diff}
\end{equation}
where $A_{4}$, $A_{2}$ and $\lambda_{c}$ are positive constants,
and $T_{P}$ is the non-dimensional temperature above which $f_{\rm
diff}$ is convex in~$c$. In the context of steel, $T_{P}$ is the
temperature above which austenite is the stable phase.  For $T < T_{P}$,
$f_{\rm diff}$ becomes a double-well function.  This can be seen in
Figure~\ref{fig:diff_free_energy}, in which the diffusive free-energy
as a function of $c$ is plotted for various temperatures.
The further the temperature drops below $T_{P}$, the deeper the
energy wells. The existence of two wells is what can lead to regions
with layers alternating values of $c$, which is typical of a pearlitic
structure. The presence of the term $\nabla c$ in the free-energy accounts
for the energy associated with the formation a phase boundaries. In the
context of pearlite, it reflects the surface energy associated with the
boundaries between cementite and ferrite.  Mathematically, it provides
a regularising effect when the free-energy is non-convex with respect
to~$c$. Note that below $T_{P}$ there is no stable local minimum at $c=0$
for the chosen free-energy function.
\subsubsection{Displacive part of the free-energy}
\label{sec:displ}
\begin{figure}
  \center\includegraphics[scale=0.6]{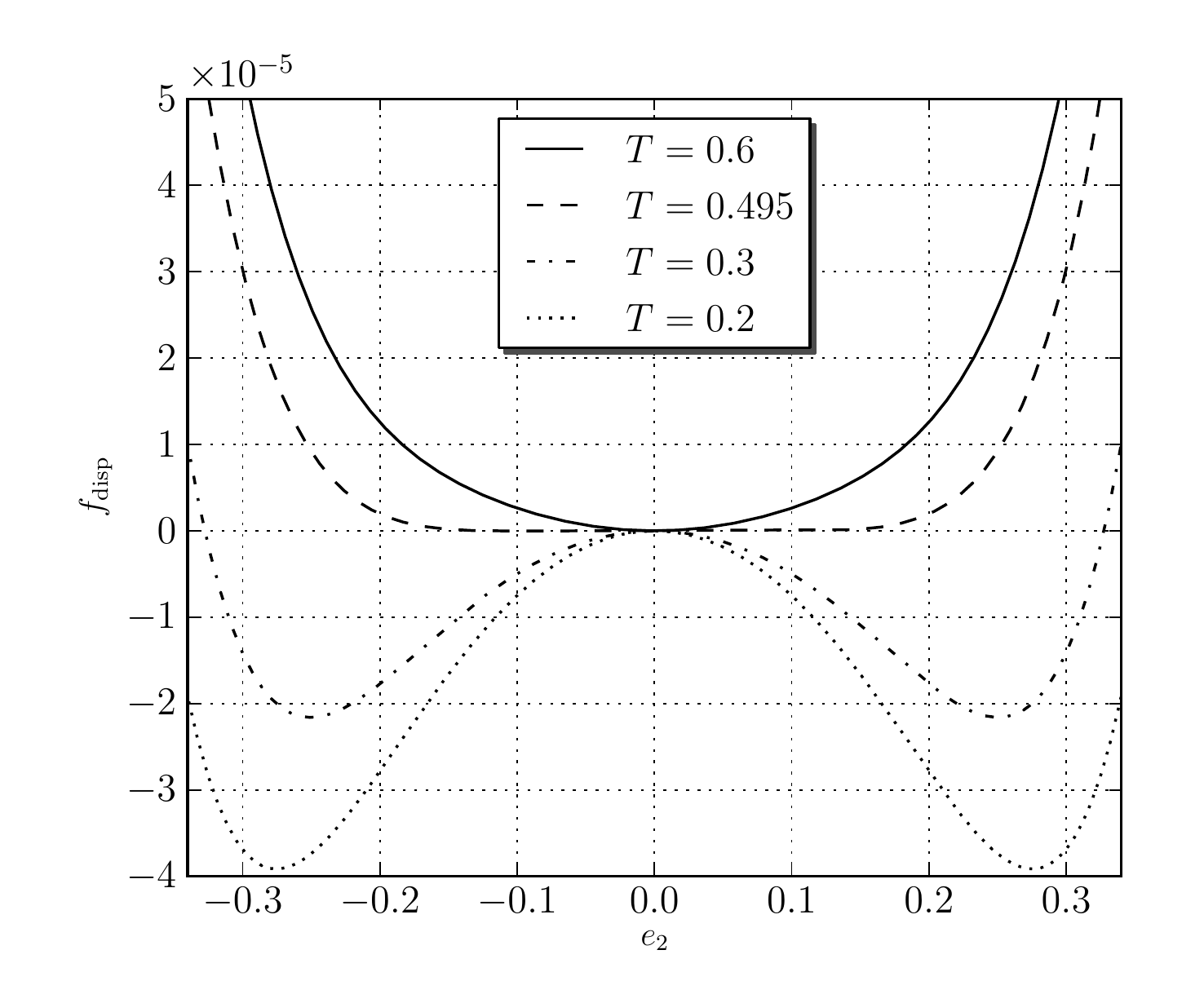}
\caption{Displacive free-energy density as a function of $e_{2}$ at various
temperatures $T$ and $e_{1} = e_{3} = 0$
($B_{62} = 3.69$, $B_{42} = 1.24 \times 10^{-1}$, $B_{22} = 4.97 \times 10^{-3}$, $T_{M}=0.495$).}
\label{fig:disp_free_energy}
\end{figure}

The displacive part of the free-energy is postulated as follows:
\begin{multline}
  f_{\rm disp}
  = \dfrac{B_{62}}{6} e_{2}^{6} - \dfrac{B_{42}}{4} e_{2}^4
      +  \dfrac{B_{22}}{2} \dfrac{T-T_M}{T_M} e_{2}^2
      +  \dfrac{B_{1}}{2}e_{1} \bracs{e_{1} - \brac{
            \alpha \brac{T-T_{\rm ref}} + x_{1c}c + x_{12} e_{2}^2 }}
\\
      +  \dfrac{B_{3}}{2} e_{3}^2
+ \frac{\lambda_e}{2}  \| \nabla e_{2} \|^{2},
\label{eqn:free_energy_displ}
\end{multline}
where $B_{62}$, $B_{42}$, $B_{22}$, $B_{1}$, $B_{3}$ and $\lambda_e$
are positive constants, $T_{M}$ is the non-dimensional temperature below
which $f_{\rm disp}$ is non-convex in~$e_{2}$, $x_{1c}$ and $x_{12}$
are constant coupling parameters that induce volumetric changes as a
consequence of diffusive and displacive phase changes, respectively,
and $\alpha > 0$ is the thermoelastic coefficient which determines
volumetric changes as a consequence of temperature deviations away from
a reference temperature $T_{\rm ref}$.  The volumetric term in the
displacive free-energy is chosen such that the model will coincide
with classical thermoelasticity.  The displacive contribution to
the free-energy as a function of $e_{2}$ at various temperatures is
illustrated in Figure~\ref{fig:disp_free_energy}.
Physically, $T_{M}$ is the temperature below which martensite can
form.  The further the temperature drops below $T_{M}$, the deeper the
double-wells in $f_{\rm disp}$ as a function of~$e_{2}$.  This is what
leads to formation of martensite, with spatially alternating values of
$e_{2}$ indicating twinning.  Analogous to the diffusive contribution to
the free-energy, the gradient term $\nabla e_{2}$ accounts for the energy
associated with twin boundaries, and provides a regularising effect
at temperatures below~$T_{M}$.  Similar to the diffusive free-energy
contribution, below $T_{M}$ there is no stable local minimum at $e_{2}=0$
for the chosen free-energy function.
\subsubsection{Thermal part of the free-energy}
\label{sec:thermal}
The thermal part of the free-energy is postulated to be of the classical
form
\begin{equation}
  f_{\rm therm}
  =
- s_{0}\brac{T-T_{\rm ref}}
- \frac{c_{v}}{2T_{\rm ref}} \brac{T-T_{\rm ref}}^{2},
\label{eqn:thermal_free_energy}
\end{equation}
where $s_{0}$ is the reference entropy density and $c_{v}$ is the heat
capacity, both of which are assumed to be positive and constant, and
$T_{\rm ref}$ is the aforementioned constant reference temperature.

\subsubsection{Phase coupling contribution}
\label{sec:coupling_energy}

Displacive and diffusive phases do not generally co-exists at a given
point.  To model this, we include in the free-energy a contribution of
the form
\begin{equation}
  f_{\rm cpl} = x_{2c} c^2 e_{2}^2,
\end{equation}
where $x_{2c} > 0$ is a constant.  The task of this term is to
penalise energetically concurrent variations away from zero of $e_{2}$
and~$c$. That is, it penalises the co-existence of pearlitic ($c \ne 0$)
and martensitic phases ($e_{2} \ne 0$).
\subsection{Constitutive models as a consequence of the free-energy}

The thermodynamic restrictions in
Section~\ref{sec:admissible_constitutive_equations}, together with
the Helmholtz free-energy defined in this section, provide constitutive
models for the stress, the chemical potential and the entropy. We provide
now some expansions for these constitutive models in terms of the order
parameters for the particular free-energy that we consider.

Introducing the notation $\Bar{\vect{\sigma}} = \pd f/ \pd
\vect{\eps}$, for the `local' part of the inviscid stress tensor (see
equation~\eqref{eqn:stress_definition}), $\Bar{\vect{\sigma}}$ in terms
of derivatives of $f$ with respect to the order parameters reads:
\begin{equation}
  \Bar{\vect{\sigma}} =
    \frac{\pd f}{\pd e_{1}}  \brac{\vect{e}_{1} \otimes \vect{e}_{1} + \vect{e}_{2} \otimes \vect{e}_{2} }
    + \frac{\pd f}{\pd e_{2}} \brac{\vect{e}_{1} \otimes \vect{e}_{1} - \vect{e}_{2} \otimes \vect{e}_{2} }
    +  \frac{\pd f}{\pd e_{3}} e_{3}\brac{\vect{e}_{1} \otimes \vect{e}_{2} + \vect{e}_{2} \otimes \vect{e}_{1} },
\end{equation}
where $\vect{e}_{i}$ is a canonical unit basis vector.  The divergence
of the higher-order stress term $\vect{\Sigma} = \pd f / \pd
\nabla\vect{\eps}$ (see equation~\eqref{eqn:Sigma_definition}) in terms
of the order parameters reads:
\begin{equation}
  \nabla \cdot \vect{\Sigma}
  =
    \nabla \cdot \brac{\frac{\pd f}{\pd \nabla e_{2}}}
      \brac{\vect{e}_{1} \otimes \vect{e}_{1} - \vect{e}_{2} \otimes \vect{e}_{2} }
  = \lambda_e \nabla^{2} e_{2}
              \brac{\vect{e}_{1} \otimes \vect{e}_{1} - \vect{e}_{2} \otimes \vect{e}_{2} }.
\label{eqn:gradient_stress}
\end{equation}
The chemical potential $\mu$ is of the form (see
equation~\eqref{eqn:mu_definition}):
\begin{equation}
  \mu = \frac{\pd f}{\pd c} - \lambda_{c} \nabla^2 c,
\end{equation}
where $\pd f / \pd c$ is the `local' part of the chemical potential.
The time derivative of the entropy density, which will play a role in
formulating the heat transport equation, reads:
\begin{equation}
\begin{split}
  \dot{s} &= - \frac{\pd^{2} f}{\pd T^{2}} \Dot{T}
             - \frac{\pd^{2} f}{\pd T \pd \vect{\eps}} : \Dot{\vect{\eps}}
             - \frac{\pd^{2} f}{\pd T \pd \nabla \vect \eps} \vdots \nabla \Dot{\vect{\eps}}
             - \frac{\pd^{2} f}{\pd T \pd c}  \dot{c}
             - \frac{\pd^{2} f}{\pd T \pd \nabla c} \cdot \nabla \Dot{c}
\\
          &=   \frac{c_{v}}{T_{\rm ref}} \Dot{T}
             - \frac{\pd^{2} f}{\pd T\pd e_{1}} \Dot{e}_{1}
             - \frac{\pd^{2} f}{\pd T\pd e_{2}} \Dot{e}_{2}
             - \frac{\pd^{2} f}{\pd T\pd c} \Dot{c},
\end{split}
\label{eqn:entropy_rate}
\end{equation}
where in the last line we have taken into account that temperature does
not affect the phase boundary energy.
\subsection{Constitutive models for the dissipative terms}

Constitutive models for the mass flux, the heat flux and the
viscous stress do not follow as a direct consequence of the chosen
of the free-energy, but must be chosen such that the restrictions
in Section~\ref{sec:admissible_constitutive_equations} are satisfied.
As is usual, the definition of a free-energy does not provide information
on the kinetics (rate) of phase transformation processes.

We opt for a viscous stress that depends on the rate of $e_{2}$ only,
\begin{equation}
  \vect{\sigma}_{v}
    = \nu \Dot{e}_{2} \brac{\vect{e}_{1} \otimes \vect{e}_{1} - \vect{e}_{2} \otimes \vect{e}_{2}},
\end{equation}
where $\nu \ge 0$, in which case
equation~\eqref{eqn:viscous_stress_restriction} is satisfied.  The mass
flux $\vect{j}_{c}$ is assumed to be proportional to the chemical
potential,
\begin{equation}
  \vect{j}_{c} = - M \nabla \mu,
\label{eqn:mass-flux}
\end{equation}
where $M$ is known as the `mobility'. The expression
\begin{equation}
  M  = M_{0}\exp\brac{-\frac{Q}{T}},
\label{eqn:mobility}
\end{equation}
where $M_{0} \ge 0$ and $Q \ge 0$ are constants, will be
used in the simulations.  The thermodynamic restriction in
equation~\eqref{eqn:mass_flux_restriction} is satisfied if~$M \ge 0$.
For the heat flux, the Fourier model is assumed:
\begin{equation}
  \vect{q} = - k \nabla T,
\end{equation}
where $k$ is the thermal conductivity. The thermodynamic restriction in
equation~\eqref{eqn:heat_flux_restriction} is satisfied if~$k \ge 0$.

\section{Governing initial boundary value problems}
\label{sect:IBVP}

A complete set of governing equations for the considered model can now
be defined. The model is based on momentum balance, mass balance for
the solute phase and an energy balance for heat transport. The primal
unknowns to be solved for are the displacement, the solute concentration
and the temperature.

The boundary $\partial \Omega$ is assumed to be partitioned into
subsets such that $\overline{\pd \Gamma_{h} \cup \pd \Gamma_{g}} =
\overline{\pd \Gamma_{q} \cup \pd \Gamma_{T}} = \pd \Omega$ and $\pd
\Gamma_{h} \cap \pd \Gamma_{g} = \pd \Gamma_{q} \cap \pd \Gamma_{T}
= \emptyset$. The balance of momentum equation follows directly from
equation~\eqref{eqn:momentum_balance_local}. Neglecting differences in
mass density between phases, the balance of momentum equation and the
associated boundary and initial conditions read:
\begin{align}
  \rho \Ddot{\vect{u}} - \nabla \cdot \vect{\sigma}
        &= \vect{b} \quad \ \ {\rm in} \ \Omega \times \bracs{0, t_{\rm f}}
\label{eqn:linear_momentum_balance}
\\
  \vect{\sigma} \vect{n} &= \vect{h}_s \quad {\rm on} \ \Gamma_{h} \times \bracs{0, t_{\rm f}}
\\
  \vect{u}               &= \vect{g}_u \quad {\rm on} \ \Gamma_{g} \times \bracs{0, t_{\rm f}}
\\
  \lambda_{e} \nabla e_{2} \cdot \vect{n} &= 0 \quad \ \ {\rm on} \ \pd\Omega \times \bracs{0, t_{\rm f}}
\label{eqn:momentum_balance-grad-bc}
\\
  \vect{u}\brac{\vect{x}, 0} &= \vect{u}_{0} \quad {\rm on} \ \Omega
\\
  \Dot{\vect{u}}\brac{\vect{x}, 0} &= \vect{v}_{0} \quad \, {\rm on} \ \Omega.
\end{align}
The boundary condition $\nabla e_{2} \cdot \vect{n} = 0$ is motivated
by the interpretation that the gradient terms provide a nonlocal effect,
and in the absence of neighbouring material particles on the boundary no
nonlocal work can be done on $\pd \Omega$.  This coined the `insulation
condition' by \citet{polizzotto:2003}.

Inserting the expression for the mass flux in
equation~\eqref{eqn:mass-flux} into the differential expression of mass
conservation~\eqref{eqn:mass_balance_integral} leads to an equation
governing the transport of the solute.  For the form of the free-energy
given in Section~\ref{sec:Helmholtz-free-energy} and the definition
of $\mu$ in equation~\eqref{eqn:mu_definition} in terms of the
free-energy, the mass diffusion equation and associated boundary and initial
conditions read:
\begin{align}
  \Dot{c} - \nabla \cdot M \nabla\brac{\frac{\pd f}{\pd c}  - \lambda \nabla^{2} c} &= 0
      \quad {\rm on} \ \Omega \times \bracs{0, t_{\rm f}}
\label{eqn:ch}
\\
  M \nabla\brac{\frac{\pd f}{\pd c}  - \lambda \nabla^{2} c} \cdot \vect{n}
     &= 0 \quad {\rm on} \ \pd\Omega \times \bracs{0, t_{\rm f}}
\label{eqn:cahn-hilliard-flux_bc}
\\
  \lambda_c \nabla c \cdot \vect{n}  &= 0 \quad {\rm on} \ \pd\Omega \times \bracs{0, t_{\rm f}}
\label{eqn:cahn-hilliard-gradient_bc}
\\
  c\brac{\vect{x}, 0} &= c_{0}  \ \ {\rm on} \ \Omega.
\end{align}
The above equation is known as the Cahn-Hilliard equation.
A consequence of the manner in which the surface energy is included in
the model is the presence of fourth-order derivatives.  The boundary
condition in~\eqref{eqn:cahn-hilliard-flux_bc} implies that there
is no mass flux across $\pd\Omega$ (see \eqref{eqn:mu_definition}
and \eqref{eqn:mass-flux}), and \eqref{eqn:cahn-hilliard-gradient_bc}
is interpreted in the same fashion as the condition on $\nabla e_{2}
\cdot \vect{n}$ in the momentum balance.  A more general form of the
boundary conditions for the Cahn-Hilliard equation can be found in
\citet{wells:2006}.

The final governing equation is the energy balance.  Using the
definition of the Helmholtz free-energy, $\Dot{u}$ can be
replaced by $\dot{f} + \dot{T}s + T\dot{s}$ in the energy
balance~\eqref{eqn:internal_energy_balance}, yielding
\begin{equation}
\dot{f} + \dot{T}s + T\dot{s}
  =
   - \nabla \cdot \vect{q}
   + \vect{\sigma} : \nabla \vect{v}
   + (\nabla \cdot \vect{\Sigma}) : \Dot{\vect{\eps}}
   + \vect{\Sigma} : \nabla \Dot{\vect{\eps}}
\\
   + (\nabla \cdot \vect{\xi}) \Dot{c}
   + \vect{\xi} \cdot \nabla \Dot{c}
   - \mu \nabla \cdot \vect{j}_{c}
   - \nabla \mu \cdot \vect{j}_{c}.
\end{equation}
By considering the expansion of~$\Dot{f}$ in~\eqref{eqn:free_energy_rate}
and the constitutive models for $\vect{\sigma}_{e}$, $\mu$ and $s$
in equations~\eqref{eqn:stress_definition}, \eqref{eqn:mu_definition}
and~\eqref{eqn:entropy_definition}, respectively, the energy balance
reduces to
\begin{equation}
T\dot{s}
  = - \nabla \cdot \vect{q} + \vect{\sigma}_{v} : \Dot{\vect{\eps}}
   - \nabla \mu \cdot \vect{j}_{c}.
\label{eqn:Ts}
\end{equation}
Using the expression of $\dot{s}$ in equation~\eqref{eqn:entropy_rate},
the heat equation, together with suitable boundary and initial conditions
reads:
\begin{align}
   \frac{T c_{v}}{T_{\rm ref}}  \Dot{T}
 - T \brac{\frac{\pd \vect{\sigma}}{\pd T}: \Dot{\vect{\eps}} + \frac{\pd \mu}{\pd T}  \Dot{c}}
 ‍ + \nabla \mu \cdot \vect{j}_{c}
 ‍ - \vect{\sigma}_{v} : \Dot{\vect{\eps}}
   + \nabla \cdot \vect{q}
  &= 0 \quad \ \ {\rm on} \ \Omega \times \bracs{0, t_{\rm f}}
\label{eqn:heat_equation}
\\
  \vect{q} \cdot \vect{n}   &= h_{T} \quad  {\rm on} \ \Gamma_{q} \times \bracs{0, t_{\rm f}}
\\
  T  &= g_{T}  \quad  {\rm on} \ \Gamma_{T} \times \bracs{0, t_{\rm f}}
\\
  T\brac{\vect{x}, 0} &= T_{0} \quad  \, {\rm on} \ \Omega.
\end{align}
The complete model involves the solution of all three coupled equations.

\section{Fully-discrete Galerkin formulation}

A Galerkin finite element formulation of the governing equations is now
developed which will be used in computing numerical examples.  A feature
of the governing equations is the presence of fourth-order derivatives of
the concentration and displacement fields. Ordinarily, this would require
finding approximate solutions in a subspace of~$H^{2}\brac{\Omega}$, but
such finite element element spaces are troublesome to construct.  We will
exploit two different strategies to avoid this difficulty. For the mass
diffusion equation it is convenient to adopt a mixed formulation, which
is essentially an operating-splitting approach and is not encumbered
with the difficulties that plague mixed formulations that result
in saddle-point problems.  For the momentum balance, mixed schemes
are difficult to analyse and in some cases stable schemes may not
be known. To circumvent this difficulty, techniques for fourth-order
elliptic equations that are inspired by discontinuous Galerkin methods
will be used (see \citet{engel:2002,wells:2007}). These methods permit
the rigorous solution of fourth-order problems using a primal formulation
and $H^{1}\brac{\Omega}$-conforming finite element spaces.

To formulate a finite element problem, let $\mathcal{T}$ be a
triangulation of $\Omega$ into finite element cells such that $\mathcal{T}
= \bracc{K}$. We will work with the usual Lagrange finite element basis
\begin{equation}
  V_{k} = \bracc{v \in H^{1}\brac{\Omega}, v \in P_{k}\brac{K} \forall K \in \mathcal{T}},
\end{equation}
where $P_{k}$ denotes the space of Lagrange polynomials of order $k$
on a finite element cell. The numerical formulation that we propose
involves solving the following problem: given the data at time $t_{n}$,
find $\vect{u}_{h}, c_{h}, \tau_{h}, T_{h} \in (V_{k_{2}})^{d} \times
V_{k_{1}} \times V_{k_{1}} \times V_{k_{1}}$ at time $t_{n+1}$ such that
\begin{equation}
  L(\vect{w}, q, r, z; \vect{u}_{h}, c_{h}, \tau_{h}, T_{h})
      = 0 \quad \forall \brac{\vect{w}, q, r, z}
      \in (V_{k_{2}})^{d} \times V_{k_{1}} \times V_{k_{1}} \times V_{k_{1}},
\label{eqn:L_functional}
\end{equation}
where $k_{2} > 1$ and $k_{1} > 0$.  The functional $L$ is linear in each
of $\brac{\vect{v}, q, r, w}$ and therefore can be split additively into
four contributions:
\begin{multline}
  L(\vect{w}, q, r, z; \vect{u}_{h}, c_{h}, \tau_{h}, T_{h})
=
    L_{u}(\vect{w}; \vect{u}_{h}, c_{h}, \tau_{h}, T_{h})
  + L_{c}(q; \vect{u}_{h}, c_{h}, \tau_{h}, T_{h})
\\
  + L_{\tau}(r; \vect{u}_{h}, c_{h}, \tau_{h}, T_{h})
  + L_{T}(z; \vect{u}_{h}, c_{h}, \tau_{h}, T_{h}),
\end{multline}
where $L_{u}$, $L_{c}$, $L_{\tau}$ and $L_{T}$ will represent the
contributions of linear momentum, mass diffusion, the chemical potential
and heat transport, respectively. Each of these terms will be defined
in this section.

Implicit time integrators will be used for all equations. We believe
this to be the only feasible approach owing to the parabolic nature of
the mass and heat transport equations, and the presence of fourth-order
terms in the momentum balance and mass diffusion equations.

It is particularly convenient to express the problem using the format
of equation~\eqref{eqn:L_functional} since we use high-level tools
to generate computer code automatically for this problem, and these
tools inherit this mathematical expressiveness.  Furthermore, the
tools perform automatic differentiation (both regular and directional
derivatives) so there is no need to compute by hand a linearisation of
the problem.  The functional $L$ will be the input to the code, and the
directional derivative (the linearisation) is computed using automatic
differentiation, yielding the Jacobian for use in a Newton solver.
This will be expanded upon in the following section.

The subscript `$h$' will be used in this section to denote an approximate
quantity, for example $f_{h} = f(\vect{u}_{h})$, where $\vect{u}_{h}$
is the approximate displacement field.

\subsection{Balance of momentum}
Multiplying the balance of momentum
equation~\eqref{eqn:linear_momentum_balance} by a function $\vect{w}$
and applying integration by parts to various terms,
\begin{multline}
  \int_{\Omega} \rho \vect{w} \cdot \Ddot{\vect{u}}_{h} \dif x
  + \int_{\Omega} \nabla \vect{w} : \Bar{\vect{\sigma}}_{h} \dif x
  + \int_{\Omega} \nabla \vect{w} : \nabla \cdot \vect{\Sigma}_{h} \dif x
  + \int_{\Omega} \nabla \vect{w} : \vect{\sigma}_{v, h} \dif x
\\
  - \int_{\Omega} \vect{w} \cdot \vect{b}  \dif x
  - \int_{\pd\Omega} \vect{w} \cdot \vect{h}_s  \dif s
= 0,
\end{multline}
where the decomposition of the stress into `local' and
`gradient' contributions has been used, $\vect{\sigma}_{h, e} =
\Bar{\vect{\sigma}}_{h} - \nabla \cdot \vect{\Sigma}_{h}$, and the
Neumann boundary condition $\vect{\sigma} \vect{n} = \vect{h}_s$ has
been inserted. Addressing now the term involving $\vect{\Sigma}_{h}$
and considering the precise form of $\vect{\Sigma}_{h}$
in~\eqref{eqn:gradient_stress}, we have
\begin{equation}
  \int_{\Omega} \nabla \vect{w} : \nabla \cdot \vect{\Sigma}_{h} \dif x
 = - \int_{\Omega} \brac{w_{1,1} - w_{2,2}} \lambda_{e} \nabla^{2} e_{2, h} \dif x.
\end{equation}
This term is problematic since after the application of integration by
parts, second-order spatial derivatives of $\vect{w}$ and $\vect{u}_{h}$
are present, which classically would require searching for approximate
solutions using $H^{2}$-conforming functions, thereby precluding
the use of standard Lagrange finite element basis functions.
To circumvent this difficulty while still using $H^{1}$-conforming
element and without abandoning a primal approach, we introduce
integrals over finite element cell facets to impose weak continuity of
the normal derivative while preserving consistency and stability, as
formulated for the biharmonic equation in \citet{engel:2002} and the
Cahn-Hilliard equation in \citet{wells:2006}.  This then permits the
use of standard $H^{1}$-conforming finite element basis functions. The
modified variational form, using $\pd f/ \pd e_{i}$ in place of
$\Bar{\vect{\sigma}}$, reads:
\begin{multline}
  L_{u} = \int_{\Omega} \rho \vect{w} \cdot \Ddot{\vect{u}}_{h, n + 1 - \alpha_{m}} \dif x
  + \int_{\Omega} (w_{1,1} + w_{2,2}) \frac{\pd f_{h, n+1-\alpha_{f}}}{\pd e_{1}} \dif x
  + \int_{\Omega} (w_{1,1} - w_{2,2}) \frac{\pd f_{h, n+1-\alpha_{f}}}{\pd e_{2}} \dif x
\\
  + \int_{\Omega} (w_{1,2} + w_{2,1}) \frac{\pd f_{h, n+1-\alpha_{f}}}{\pd e_{3}} \dif x
  + \sum_{K} \int_{K} \nabla \brac{w_{1,1} - w_{2,2}} \cdot \lambda_{e} \nabla e_{2, h, n+1-\alpha_{f}} \dif x
\\
  - \int_{\Tilde{\Gamma}} \jump{w_{1,1} - w_{2,2}} \cdot \avg{\lambda_{e} \nabla e_{2, h, n+1-\alpha_{f}}} \dif s
  - \int_{\Tilde{\Gamma}} \avg{\lambda_{e} \nabla \brac{w_{1,1} - w_{2,2}}} \cdot \jump{e_{2, h, n+1-\alpha_{f}}} \dif s
\\
  + \int_{\Tilde{\Gamma}} \frac{\eta \lambda_{e}}{h_{K}}  \jump{w_{1,1} - w_{2,2}} \cdot \jump{e_{2, h, n+1-\alpha_{f}}} \dif\Gamma
  + \int_{\Omega} (w_{1,1} - w_{2,2}) \nu e_{2, h, n+1-\alpha_{f}}\dif x
\\
  - \int_{\Omega} \vect{w} \cdot \vect{b}_{n+1-\alpha_{f}}  \dif x
  - \int_{\pd\Omega} \vect{w} \cdot \vect{h}_{s, {n+1-\alpha_{f}}}  \dif s,
\end{multline}
where $\Tilde{\Gamma}$ denotes the set of all interior facets, $\jump{a} =
a_{+} \vect{n}_{+} + a_{-} \vect{n}_{-}$ and $\avg{b} = \brac{\vect{b}_{+}
+ \vect{b}_{-}}/2$, `$+$' and `$-$' denote opposite sides of a cell
facet, $\eta$ is a dimensionless penalty parameter and $h_{K}$ is a
measure of the local element size. The penalty parameter is required
for stability of the formulation and is of order one.  Time derivatives
are dealt with using a generalised-$\alpha$ scheme (see, for example,
\citet{chung:1993}). The acceleration, velocity and displacement at the
end of a time step are related via
\begin{align}
  \vect{u}_{n+1}
    &= \vect{u}_{n} + \Delta t \Dot{\vect{u}}_{n}
         + \Delta t^{2}\brac{ \brac{\frac{1}{2} - \beta}\Ddot{\vect{u}}_{n} + \beta\Ddot{\vect{u}}_{n+1}},
\\
  \Dot{\vect{u}}_{n+1}
    &= \Dot{\vect{u}}_{n} + \Delta t\brac{\brac{1-\gamma}\Ddot{\vect{u}}_{n} + \gamma\Ddot{\vect{u}}_{n+1}},
\end{align}
where $\Delta t = t_{n+1} - t_{n}$, $\beta$ and $\gamma$ are parameters,
and mid-point values of primal fields are computed according to
\begin{equation}
  y_{n+1-\alpha} = \brac{1- \alpha}y_{n+1} + \alpha y_{n}.
\end{equation}
where $\alpha$ is a parameter. The parameters $\alpha_{m}$, $\alpha_{f}$,
$\gamma$ and $\beta$ determine properties of the time stepping scheme and
will be reported with other problem data for the numerical examples in
the following section.  Nonlinear functions are evaluated at a mid-point,
for example
\begin{equation}
  f_{n+\alpha} = f(T_{n+\alpha}, e_{i,n+\alpha}, \nabla e_{2,n+\alpha}, c_{n+\alpha}, \nabla c_{n+\alpha}).
\end{equation}

The discontinuous Galerkin-type approach is consistent for $k > 1$, and
analysis of the biharmonic equation has shown that the method is stable
for sufficiently large~$\alpha$ (typically $\alpha$ is of order~$1$).
It has been proven that the method converges in the $L^{2}$ for the
biharmonic equation at a rate of $k+1$ for $k > 2$~\citep{engel:2002}
and at a rate of $k$ for $k = 2$~\citep{wells:2007}.
\subsection{Mixed form of the Cahn-Hilliard equation}
For the Cahn-Hilliard equation we adopt an operator splitting
approach to deal with the fourth-order spatial derivative and split
equation~\eqref{eqn:ch} into two second-order equations:
\begin{align}
  \Dot{c} -\nabla \cdot M \nabla \tau &= 0,
\label{eqn:CH-mixed-mass-term}
\\
  \tau - \frac{\pd f}{\pd c} + \lambda_{c} \nabla^{2} c &= 0,
\label{eqn:CH-mixed-chemical-term}
\end{align}
where $c$ and $\tau$ are the independent unknowns that will be solved
for. If the above problem is solved exactly, then $\tau$ and $\mu$ will
coincide (see equation~\eqref{eqn:mu_definition}). It will be important to
distinguish between $\tau$ and $\mu$ when finding approximate solutions.
Note that through $\pd f/\pd c$ there is a dependency on the strain and
the temperature.

Casting the above set of equations into a weak form, using the
Crank-Nicolson method for the time derivative, and applying the boundary
conditions from equations~\eqref{eqn:cahn-hilliard-gradient_bc}
and~\eqref{eqn:cahn-hilliard-flux_bc}, the functionals $L_{c}$ and
$L_{\tau}$ read:
\begin{align}
  L_{c} &= \int_{\Omega} q \frac{c_{h,n+1} - c_{h,n}}{\Delta t} \dif x
              + \int_{\Omega} \nabla q \cdot M \nabla \tau_{h,n+1/2} \dif x,
\\
  L_{\mu} &= \int_{\Omega} r \tau_{h, n+1} \dif x
            - \int_{\Omega} r \frac{\pd f_{h,n+1}}{\pd c} \dif x
        - \int_{\Omega} \nabla r \cdot \lambda_{c} \nabla c_{h, n+1} \dif x.
\end{align}
This operator splitting approach to the Cahn-Hilliard equation was
presented and analysed by \citet{elliott:1989}, and for the form of
the free-energy and mobility considered by \citet{elliott:1989} it was
shown to be stable. A discontinuous Galerkin type approach, similar
to that used for the momentum equation, can also be used to solve the
Cahn-Hilliard equation in its primal form \citep{wells:2006}.

\subsection{Heat transport equation}
The heat equation~\eqref{eqn:heat_equation} can be cast into a weak form
via the usual process. Using the Crank-Nicolson method to deal with time
derivatives, the functional $L_{T}$ reads:
\begin{multline}
  L_{T} = \int_{\Omega} z \frac{T_{h, n+1/2}}{T_{\rm ref}} c_{v} \frac{T_{h, n+1} - T_{h,n}}{\Delta t} \dif x
\\
 - \int_{\Omega} z T_{h, n+1/2} \brac{\frac{\pd \vect{\sigma}_{e, h, {n+1/2}}}{\pd T}
      : \frac{\vect{\eps}_{h,n+1}- \vect{\eps}_{h,n}}{\Delta t}
 + \frac{\pd \mu_{n+1/2}}{\pd T} \frac{c_{h, n+1} - c_{h, n}}{\Delta t} } \dif x
\\
 ‍ + \int_{\Omega} z \brac{\nabla \tau_{h, n+1/2} \cdot \vect{j}_{n+1/2}
       - \vect{\sigma}_{v, {n+1/2}} : \frac{\vect{\eps}_{n+1}- \vect{\eps}_{n}}{\Delta t}} \dif x
\\
  -  \int_{\Omega} \nabla z \cdot \vect{q}_{h, n+1/2} \dif x
  - \int_{\Gamma_{q}} z h_{q, _{n+1/2}} \dif s.
\end{multline}
Taking into account details of the constitutive models,
\begin{multline}
  L_{T} = \int_{\Omega} z \brac{\frac{T_{h, n+1/2}}{T_{\rm ref}} c_{v} \frac{T_{h, n+1} - T_{h,n}}{\Delta t} \dif x
 - \sum _{i=1}^{3} T_{h, n+1/2} \brac{\frac{\pd^{2} f}{\pd e_{i} \pd T}
       \frac{e_{i,h,n+1}- e_{i, h,n}}{\Delta t} } }\dif x
\\
 -  \int_{\Omega} z \brac{ T_{h, n+1/2} \frac{\pd^{2} f}{\pd c \pd T}
       \frac{c_{i,h,n+1}- c_{i, h,n}}{\Delta t} + \nabla \tau_{h, n+1/2} \cdot M_{h} \nabla \tau_{h, n+1/2}  }\dif x
\\
 ‍ - \int_{\Omega} z \nu e_{2,h,n+1/2} \frac{e_{2, h, n+1}- e_{2, h, n}}{\Delta t} \dif x
  +  \int_{\Omega} \nabla z \cdot k_{T} \nabla T_{h, n+1/2} \dif x
  - \int_{\Gamma_{q}} z h_{q, _{n+1/2}} \dif s.
\end{multline}
%
\section{Numerical examples}
\label{sec:examples}

A number of numerical examples are now presented to demonstrate that
the model can qualitatively capture classical processes observed in the
heat treatment of steels. It is not the intention to consider realistic
material parameters at this stage. The determination of physical
parameters, and computing with these, is a substantial undertaking and
is the subject of ongoing work.

\begin{figure}
  \center\includegraphics[width=0.5\textwidth]{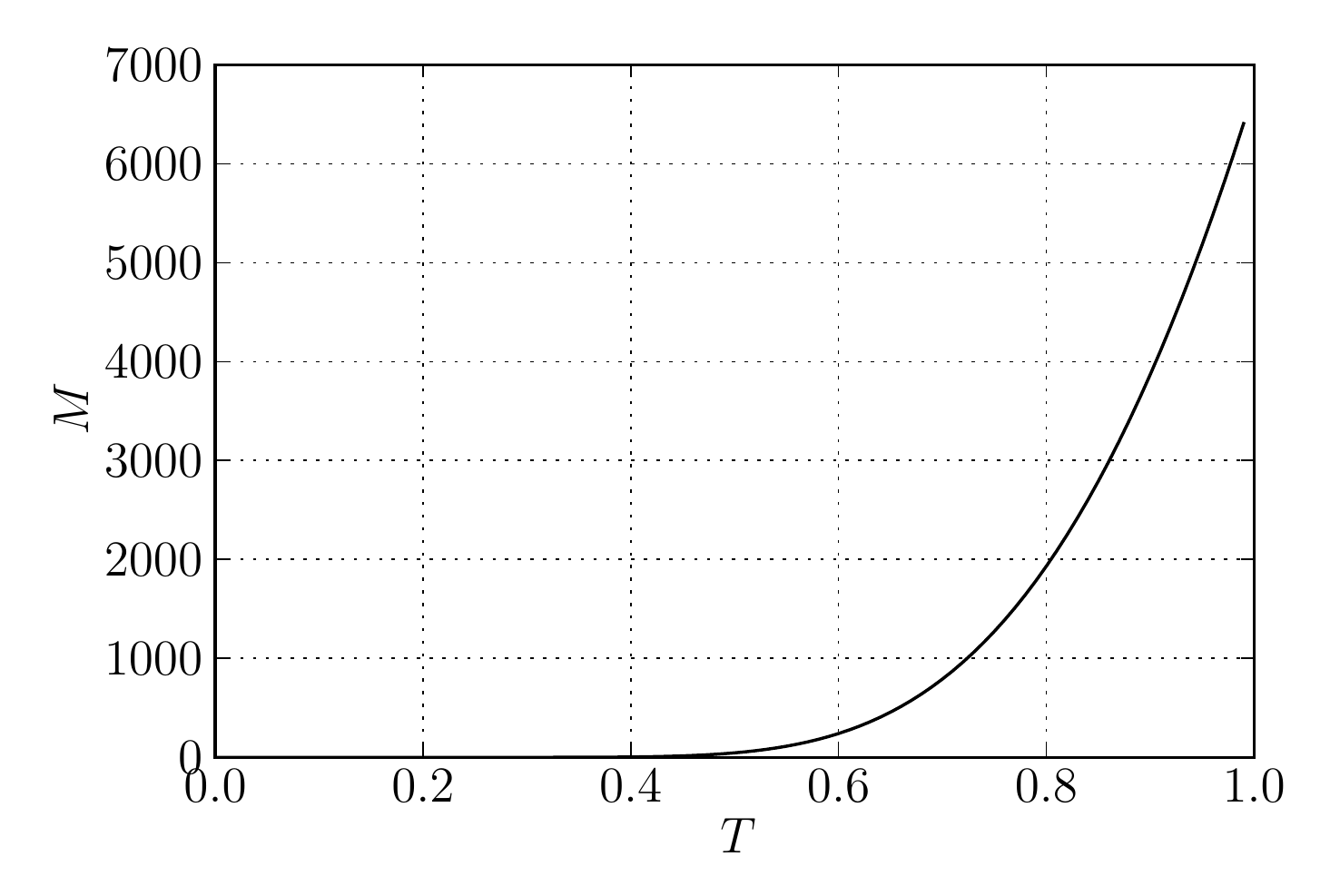}
\caption{Mobility as a function of temperature.}
\label{fig:mobility}
\end{figure}

The model includes a number of coupled processes, which permits
considerable freedom in how transformations are induced, and a
variety of parameters play a role in the development of microstructure.
The presented examples involve varying the heat flux across the boundary
of the domain for a fixed set of model parameters.  For the parameters
appearing in the free-energy, the following values are used:
$A_{4} = 7.3 \times 10^{-3}$;
$A_{2} = 6.6 \times 10^{-3}$;
$T_{P} =1$;
$\lambda_{c} = 1 \times 10^{-7}$;
$B_{62} = 0.3$;
$B_{42} = 3.1 \times 10^{-3}$;
$B_{22} = 2.5 \times 10^{-3}$;
$B_{3} = 5 \times 10^{-2}$;
$\lambda_{e}  = 1 \times 10^{-8}$;
$T_{M} = 0.495$;
$\alpha  =1 \times 10^{-2}$;
$x_{1c}  = 5 \times 10^{-3}$;
$x_{12}  = 1\times 10^{-1}$;
$c_{T}  =3.5 \times 10^{-8}$;
and $x_{2c} = 8 \times 10^{-2}$.
For model parameters that do not appear in the free-energy, the following
values are used:
$\rho = 5 \times 10^{-7}$;
$\nu  = 5 \times 10^{-8}$;
$k_{T}  = 2 \times 10^{-2}$;
$M_{0}  = 1 \times 10^{4}$; and
$Q  = 5$.
The adopted time stepping parameters are
$\rho_{\infty}  = 0.7$;
$\alpha_{m} = (2\rho_{\infty} -1)/(\rho_{\infty}+ 1)$;
$\alpha_{f} = \rho_{\infty}/(\rho_{\infty} + 1)$;
$\beta  = (1 - \alpha_{m} + \alpha_{f})^{2}/4$;
and
$\gamma  = 1/2 - \alpha_{m} + \alpha_{f}$.
The time stepping parameters are chosen such that the momentum balance
scheme is second-order accurate and strongly stable (at least for linear
problems).  The mobility $M$ as a function of temperature for the adopted
parameters is plotted in Figure~\ref{fig:mobility}.  There is considerable
scope for a extensive computational studies into the impact of various
parameters on different microstructural processes.

All simulations are performed on a unit domain $\Omega = (0,1) \times
(0,1)$ using triangular finite element cells in a regular diagonal pattern
and with 128 vertices in each direction.  For the displacement field,
$\Gamma_{g} = {(x, 0) \cup (0, y) \in \pd \Omega}$ with $\vect{u} =
\vect{0}$ on $\Gamma_{g}$ and $\vect{h}_s = \vect{0}$ on $\Gamma_{h}
= \pd \Omega \backslash \Gamma_{g}$.  With this boundary condition,
any effect of a displacement constraint on microstructure evolution
will be evident.  The heat flux across $\pd\Omega$, which we denoted by
$h_{q}$, is set proportional to the difference between the temperature
$T$ on the boundary and a prescribed `external' temperature $T_{\rm ext}$:
\begin{equation}
  h_{q} = -\delta\brac{T - T_{\rm ext}},
\end{equation}
where $\delta \ge 0$ is a parameter. The larger the value of $\delta$,
the more rapid the cooling.  Other boundary conditions were defined
in Section~\ref{sect:IBVP}.  The impact of the cooling rate on the
microstructure development will be examined by varying $\delta$
and~$T_{\rm ext}$.  The initial conditions for the displacement and
concentration are fields that are perturbed randomly about zero. In the
case of the concentration field, $c_{0} \in \bracs{-10^{-4}, 10^{-4}}$
is a random number with a uniform distribution, and for the initial
displacement, $\vect{u}_{0} \in \bracs{-10^{-4}h_{K}, 10^{-4}h_{K}}^{2}$
is a random vector with a uniform distribution and $h_{K}$ is a measure
of the finite element cell size. The dependency on the cell size is
introduced so that the magnitude of the initial strains is roughly equal
on different meshes.  In all simulations $T_{\rm ref} = T_{0} = 1.2$.

Quadratic Lagrange functions are used for the displacement
field\footnote{It is known that the formulation we have adopted to
deal with the fourth-order derivatives of the displacement field
leads to order two convergence in the $L^{2}$-norm for the biharmonic
equation when $k_{2} = 2$~\citep{wells:2007}, which may appear to be a
sub-optimal choice.  However, we are interested primarily in $e_{2}$,
and the adopted method converges with order two in the $H^{1}$-norm
when~$k_{2} = 2$.} and linear Lagrange functions are used for all other
fields ($k_{2} = 2$ and $k_{1} = 1$ in equation~\eqref{eqn:L_functional}).
A fully-coupled solution strategy is used and a Newton-Krylov method is
employed to solve the nonlinear equations in each time step.  For the
stabilising penalty term, $\eta = 8$.  The problem-specific parts of
the computer code used to perform the simulations have been generated
automatically from a high-level description that resembles closely the
notation used in this work by using a number of tools from the FEniCS
Project \citep{logg:2010,oelgaard:2010,oelgaard:2008}.  The complete
computer code used to perform all simulations reported in this work
consists of one file only and is freely available under a GNU public
license for both scrutiny and use \citep{supporting_material:2010}.

\subsection{Rapid boundary cooling to $T_{\rm ext} = 0.1$}

\begin{figure}
\center
\begin{tabular}{ccccc}
    \includegraphics[width=0.21\textwidth]{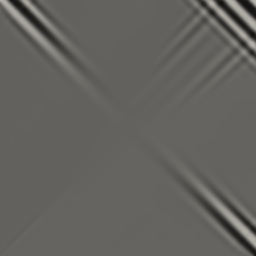}
  & \includegraphics[width=0.21\textwidth]{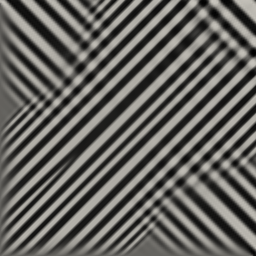}
  & \includegraphics[width=0.21\textwidth]{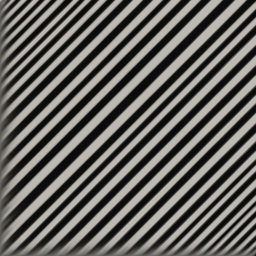}
  & \includegraphics[width=0.21\textwidth]{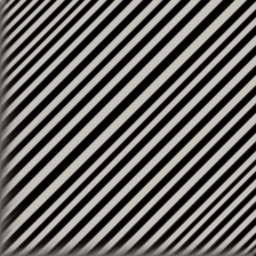}
  & \includegraphics[height=0.21\textwidth]{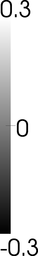}
\\
    $t = 0.01$ & $t = 0.015$ & $t = 0.025$ & $t = 0.3$ &
\end{tabular}
\caption{Contours of the displacive order parameter $e_{2}$ for the
rapid cooling case to $T_{\rm ext}= 0.1$.  The approximate range of
temperatures in the domain at the snapshots are:
at $t = 0.01$, $T \approx 0.27 - 0.61$;
at $t = 0.015$, $T \approx 0.17 - 0.31$;
at $t = 0.025$, $T \approx 0.1$; and
at $t = 0.3$, $T \approx 0.1$.}
\label{fig:rapid_0.1}
\end{figure}

We first consider rapid cooling from an initial uniform temperature of
$T=1.2$, at which temperature austenite is the stable phase, down to
$T_{\rm ext} = 0.1$, which is well-below the martensitic transition
temperature of $T_{M} = 0.495$. At $T = 0.1$, the mobility of the
diffusive phase is negligible, hence for this rapid cooling case
the formation of a diffusive phase is not anticipated; the resulting
structure will therefore be martensite. To simulate rapid cooling,
the boundary heat flux parameter $\delta = 5 \times 10^{-2}$ is used.
For this case, simulations were performed with an initial time step of
$\Delta t = 1 \times 10^{-4}$, which was increased to $\Delta t = 2 \times
10^{-3}$ at $t = 0.266$.  The computed order parameter $e_{2}$ at various
times during the cooling process is shown in Figure~\ref{fig:rapid_0.1}.
The diffusive variable $c$ (not shown) is very close to zero everywhere in
the domain. The rapid formation of fine martensite twins can be observed
in Figure~\ref{fig:rapid_0.1}.  Note that the martensite twins are
slightly less developed at the left-hand and lower edges of the domain,
due to the constraint $\vect{u} = \vect{0}$, which inhibits lattice
distortion on these constrained boundaries. This is also what induces
the bottom left-hand corner to top right-hand corner orientation of the
twins, rather than at 90$^{\circ}$ to this orientation.

\subsection{Boundary cooling to $T_{\rm ext} = 0.3$ at different rates}

\begin{figure}
\center
\begin{tabular}{ccccc}
    \includegraphics[width=0.21\textwidth]{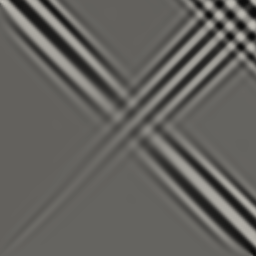}
  & \includegraphics[width=0.21\textwidth]{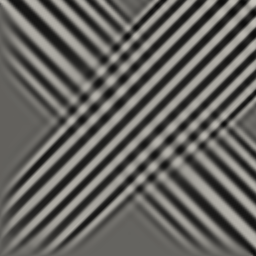}
  & \includegraphics[width=0.21\textwidth]{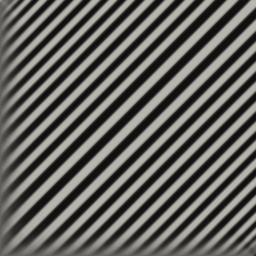}
  & \includegraphics[width=0.21\textwidth]{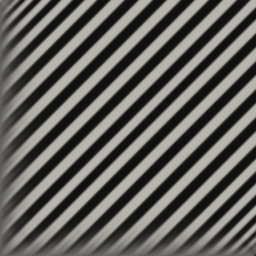}
  & \includegraphics[height=0.21\textwidth]{png/e2_legend.png}
\\
    $t = 0.0175$ & $t = 0.02$ & $t = 0.05$ & $t = 0.5$ &
\end{tabular}
\caption{Contours of the displacive order parameter $e_{2}$ for the
rapid cooling case ($\delta = 5 \times 10^{-2}$) to $T_{\rm ext}= 0.3$.
The approximate range of temperatures in the domain at the snapshots are:
at $t = 0.0175$,  $T \approx 0.35 - 0.45$;
at $t = 0.02$,    $T \approx 0.32 - 0.41$;
at $t = 0.05$,    $T \approx 0.3$;
and at $t = 0.5$, $T \approx 0.3$.}
\label{fig:rapid_0.3}
\end{figure}

We now consider a less deep quench to $T_{\rm ext} = 0.3$, which is still
below the martensitic transition temperature, with the same boundary
heat flux parameter of $\delta = 5 \times 10^{-2}$. At $T = 0.3$, the
mobility of the diffusive phase is still negligible.  For this case,
simulations were performed with a time step of $\Delta t = 2.5 \times
10^{-4}$.  The order parameter $e_{2}$ at various times during the
cooling process is shown in Figure~\ref{fig:rapid_0.3}.  For this case,
the field $c$ (not shown) is close to zero throughout the simulation,
despite the diffusive phase being energetically favourable over the
displacive phase, because diffusive processes are unable to develop due
to the brief length of time over which the temperature, and hence the
mobility, is sufficiently high.  As with the $T_{\rm ext} = 0.1$ case,
martensite forms quickly, although it is now slightly coarser in the
early stages. On a slightly longer time scale, there is some limited twin
boundary motion as part of a minimal coarsening process. Such coarsening
is not observed at lower temperatures, and can be attributed to the
different balance between bulk and surface energy (in the model, the
surface energy is unaffected by temperature while the bulk energy is).
Noteworthy is that the microstructure that forms in this case is more
affected by the displacement constraint on the two sides of the domain.
This can be explained by the smaller thermodynamic force driving the
$e_{2}$ patterning compared to the $T_{\rm ext} = 0.1$ case.

\begin{figure}
\center
\begin{tabular}{cc}
    \includegraphics[width=0.3\textwidth]{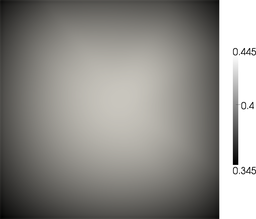}
  & \includegraphics[width=0.3\textwidth]{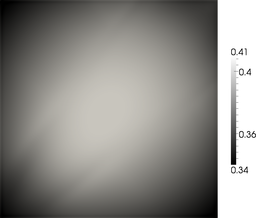}
\\
    $t = 0.0175$ & $t = 0.02$
\end{tabular}
\caption{Temperature contours for the rapid cooling case ($\delta =
5 \times 10^{-2}$) to $T_{\rm ext}= 0.3$. Thermal coupling effects
lead to the subtle breaking of symmetry in the temperature contours.
Symmetry is restored at later times via heat conduction.}
\label{fig:rapid_0.3_T}
\end{figure}

For this cooling case, the temperature contours are shown at two
snapshots in Figure~\ref{fig:rapid_0.3_T}. In the absence of thermal
coupling effects, the temperature contours would be symmetric about
the $x$- and $y$-axes.  Figure~\ref{fig:rapid_0.3_T} shows a subtle
breaking of symmetry.  In particular, at $t=0.02$ an alignment of
contours with the direction of the martensite twins can be detected.
The presented simulations are driven by large changes in the boundary
heat flux, which makes it difficult to detect local temperature changes
due to phase changes since these are relatively small, but also with
possibly high gradients that are smoothed rapidly by conduction.  The heat
capacity plays a large role in temperature changes since it determines
the change in temperature associated with a given energy release during
a transformation.

\begin{figure}
\center
\begin{tabular}{lccccc}
$e_{2}$
  & \includegraphics[width=0.19\textwidth]{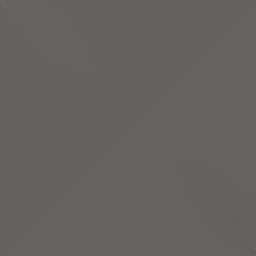}
  & \includegraphics[width=0.19\textwidth]{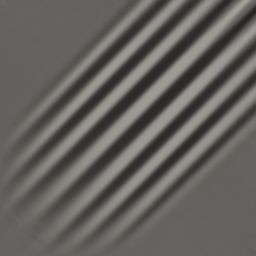}
  & \includegraphics[width=0.19\textwidth]{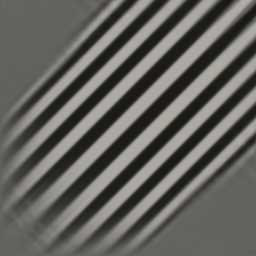}
  & \includegraphics[width=0.19\textwidth]{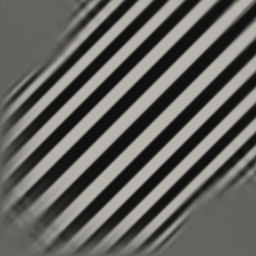}
  & \includegraphics[height=0.19\textwidth]{png/e2_legend.png}
\\
$c$
  & \includegraphics[width=0.19\textwidth]{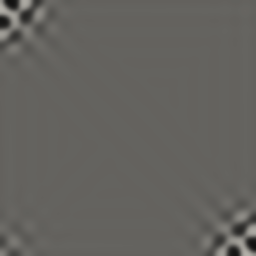}
  & \includegraphics[width=0.19\textwidth]{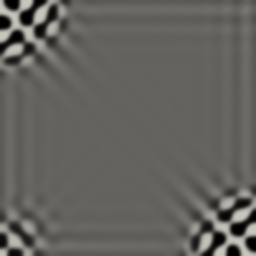}
  & \includegraphics[width=0.19\textwidth]{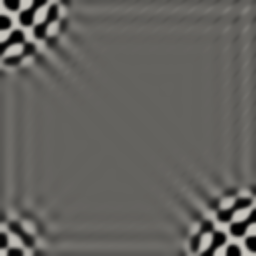}
  & \includegraphics[width=0.19\textwidth]{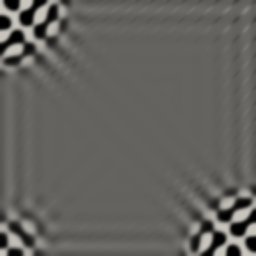}
  & \includegraphics[height=0.19\textwidth]{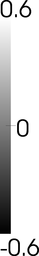}
\\
  &   $t = 0.25$ & $t = 0.4$ & $t = 0.5$ & $t = 2$ &
\end{tabular}
\caption{The displacive order parameter $e_{2}$ and the diffusive order
parameter $c$ for the moderate cooling rate case to $T_{\rm ext}= 0.3$.
The approximate temperature in the domain at the snapshots is:
at $t = 0.25$,  $T \approx 0.62$;
at $t = 0.4$,    $T \approx 0.45$;
at $t = 0.5$,    $T \approx 0.38$;
and at $t = 2$, $T \approx 0.3$.}
\label{fig:moderate_0.3}
\end{figure}

\begin{figure}
\center
\begin{tabular}{cc}
   \includegraphics[width=0.3\textwidth]{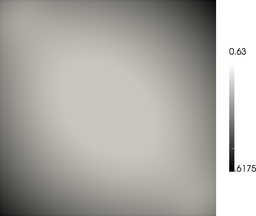}
  & \includegraphics[width=0.3\textwidth]{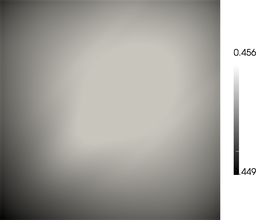}
\\
    $t = 0.25$ & $t = 0.4$
\\
   \includegraphics[width=0.3\textwidth]{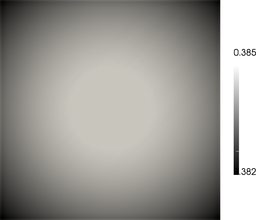}
  & \includegraphics[width=0.3\textwidth]{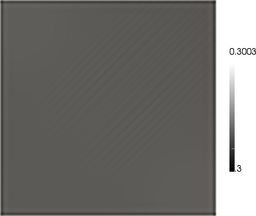}
\\
    $t = 0.5$ & $t = 2$
\end{tabular}
\caption{Temperature contours for the moderate cooling rate case
to $T_{\rm ext}= 0.3$.  Variations of the temperature contours away
from a symmetric pattern are attributable to thermal coupling effects.
At $t=2$, very subtle temperature contours aligned with the martensite
twins can be detected.}
\label{fig:moderate_0.3_T}
\end{figure}

An intermediate cooling rate case with $\delta = 1.5 \times 10^{-3}$
and a time step of $\Delta t = 2 \times 10^{-3}$ is now considered.
For this case, the order parameters $e_{2}$ and $c$ are shown in
Figure~\ref{fig:moderate_0.3} at a number if time steps.  At this cooling
rate, the impact of non-isothermal effects are clear.  There is sufficient
time for a diffusive phase to develop in three of the corners. The
formation of a diffusive phase in the corners is accelerated by the
combination thermoelastic effects and the boundary constraints and the
temperature changes.  Upon cooling, regions of volumetric straining with
opposite signs develop in three of the corners which, due to the $x_{1c}
\ne 0$, encourages diffusion of $c$. However, before a diffusive phase
can form in the entire domain, the temperature drops to a level at which
the mobility is too low for diffusion to continue, and martensite twins
form in the regions in which the diffusive phase has not yet developed.
For $t > 1$, the phase configuration is essentially stable and no further
changes could be detected.  Figure~\ref{fig:moderate_0.3_T} shows the
temperature contours for this case at four time instants. Initially a
bias can be detected with the warmest region running between the two
regions with the most developed diffusive phases (at $t=0.25$). As
time progresses, a subtle alignment of the temperature contours can be
detected. At long times, the temperature field is smoothed by conduction,
but at $t= 0.2$ very subtle temperature variations can be detected,
with the temperature contours aligned with the martensite twins. This
can be attributed to very slow twin coarsening effects.

\begin{figure}
\center
\begin{tabular}{ccccc}
    \includegraphics[width=0.21\textwidth]{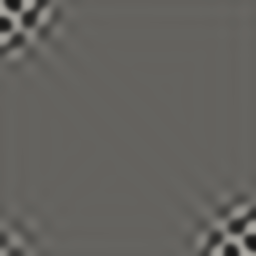}
  & \includegraphics[width=0.21\textwidth]{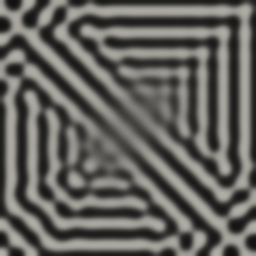}
  & \includegraphics[width=0.21\textwidth]{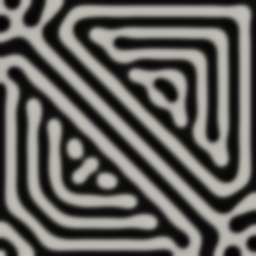}
  & \includegraphics[width=0.21\textwidth]{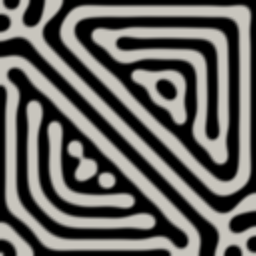}
  & \includegraphics[height=0.21\textwidth]{png/c_legend.png}
\\
    $t = 0.3$ & $t = 0.4$ & $t = 0.6$ & $t = 4$ &
\end{tabular}
\caption{Contours of the diffusive order parameter $c$ for the slow
cooling case ($\delta = 1 \times 10^{-3}$) to $T_{\rm ext}= 0.3$.
The approximate temperature in the domain at the snapshots is:
at $t = 0.3$,  $T \approx 0.72$;
at $t = 0.4$,    $T \approx 0.76$;
at $t = 0.6$,    $T \approx 0.66$;
and at $t = 4$, $T \approx 0.3$.}
\label{fig:slow_0.3}
\end{figure}

\begin{figure}
\center
\begin{tabular}{ccccc}
    \includegraphics[width=0.21\textwidth]{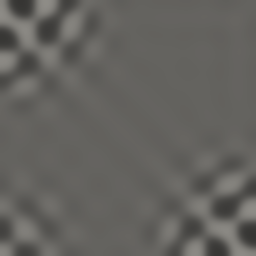}
  & \includegraphics[width=0.21\textwidth]{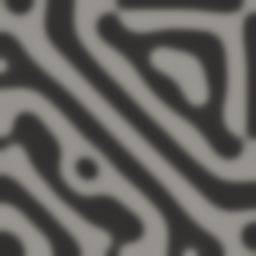}
  & \includegraphics[width=0.21\textwidth]{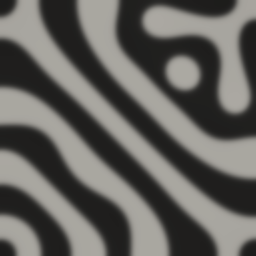}
  & \includegraphics[width=0.21\textwidth]{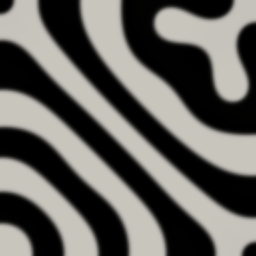}
  & \includegraphics[height=0.21\textwidth]{png/c_legend.png}
\\
    $t = 1.75$ & $t = 2.5$ & $t = 4$ & $t = 8$ &
\end{tabular}
\caption{Contours of the diffusive order parameter $c$ for the very
slow cooling case ($\delta = 1 \times 10^{-4}$) to $T_{\rm ext}= 0.3$.
The approximate temperature in the domain at the snapshots is:
at $t = 1.75$,  $T \approx 0.91$;
at $t = 2.5$,    $T \approx 0.89$;
at $t = 4$,    $T \approx 0.8$;
and at $t = 8$, $T \approx 0.6$.}
\label{fig:really_slow_0.3}
\end{figure}

Even slower cooling rate cases are now considered, from which the
impact of the cooling rate on the structure of the diffusive phase
can be observed.  For $\delta = 1 \times 10^{-3}$ and $\Delta t =
2 \times 10^{-3}$, the evolution of the diffusive phase is shown in
Figure~\ref{fig:slow_0.3}.
At this cooling rate, a reasonably fine microstructure develops,
with considerable domain boundary coherence. That is, the lamella
near the boundaries are aligned with the boundary.  Between $t=0.3$
and $t=0.4$ the temperature in the domain increases due to the energy
released during the transformation, and the time is insufficient for
conduction to transport the heat across the boundary of the domain.
The microstructural features change if the cooling rate is slowed further,
as shown in Figure~\ref{fig:really_slow_0.3}, where $\delta = 1 \times
10^{-4}$ and $\Delta t = 1 \times 10^{-2}$.
The microstructure is clearly coarser, and the boundary coherence
is reduced.  Beyond $t=8$, further microstructural changes could not
be perceived.
\subsection{Rapid boundary cooling to $T_{\rm ext} = 0.4$}

\begin{figure}
\center
\begin{tabular}{lcccccc}
$e_{2}$
  & \includegraphics[width=0.155\textwidth]{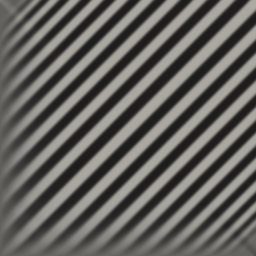}
  & \includegraphics[width=0.155\textwidth]{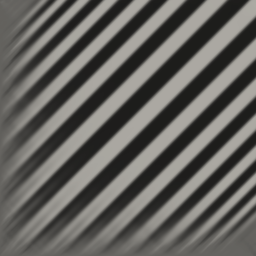}
  & \includegraphics[width=0.155\textwidth]{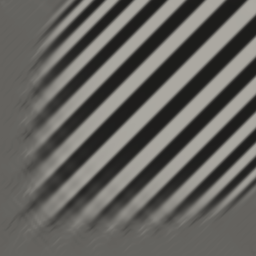}
  & \includegraphics[width=0.155\textwidth]{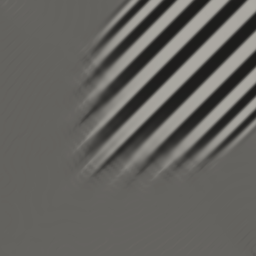}
  & \includegraphics[width=0.155\textwidth]{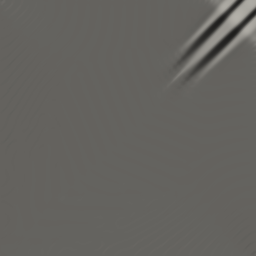}
  & \includegraphics[height=0.155\textwidth]{png/e2_legend.png}
\\
$c$
  & \includegraphics[width=0.155\textwidth]{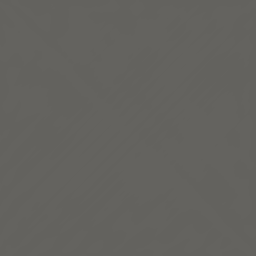}
  & \includegraphics[width=0.155\textwidth]{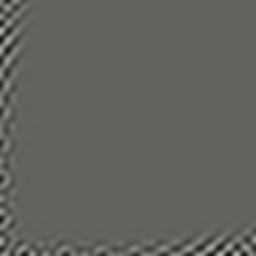}
  & \includegraphics[width=0.155\textwidth]{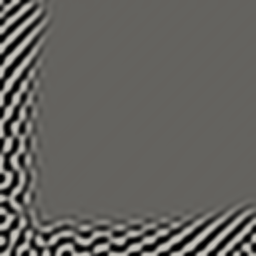}
  & \includegraphics[width=0.155\textwidth]{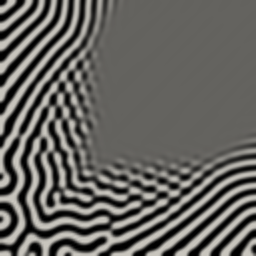}
  & \includegraphics[width=0.155\textwidth]{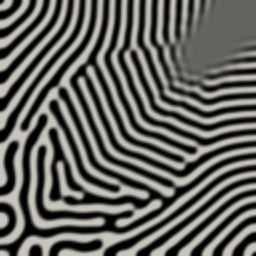}
  & \includegraphics[height=0.155\textwidth]{png/c_legend.png}
\\
  &   $t = 0.1$ & $t = 6$ & $t = 9$ & $t = 15$ & $t = 24$ &
\end{tabular}
\caption{The displacive order parameter $e_{2}$ and the diffusive order
parameter $c$ for the rapid cooling case to $T_{\rm ext}= 0.4$.  At all
snapshots, the temperature in the domain is approximately~$0.4$.}
\label{fig:rapid_0.40}
\end{figure}

The domain is now cooled to $T=0.4$ with $\delta = 5 \times 10^{-2}$.
The time step was changed during the simulation, and ranged between
$\Delta t = 2.5 \times 10^{-4}$ and $\Delta t = 5 \times 10^{-2}$.
At $T=0.4$, diffusive processes are very slow, but not yet negligible.
For this case, the computed contours of $e_{2}$ and $c$ are presented
in Figure~\ref{fig:rapid_0.40}.
It is clear that martensite twins form initially. The thermodynamic
driving force for the development of a diffusive phase is greater than
for the displacive phase, although it is inhibited by the low mobility,
but eventually diffusive process can be detected and the diffusive phase
slowly replaces the displacive phase. At this external temperature,
the structure of the diffusive phase is quite fine.

An extremely challenging aspect of computations at this temperature
is reconciling the dramatic time scale differences for the different
processes, especially during the latter stages.  The diffusive phase
develops very slowly, but the evolution of the diffusive phase can induce
sporadic changes in the martensitic phase.  Using a time step that is
suitable for diffusive processes can lead to a loss of solution stability
when sporadic displacive changes take place. There is considerable scope
for developing efficient, robust and accurate methods for spanning the
large difference in time scales.
\subsection{Rapid boundary cooling to $T_{\rm ext} = 0.45$}

\begin{figure}
\center
\begin{tabular}{lcccccc}
$e_{2}$
  & \includegraphics[width=0.15\textwidth]{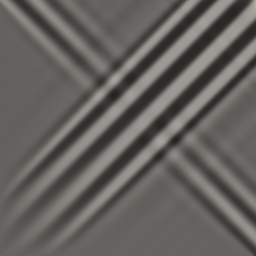}
  & \includegraphics[width=0.15\textwidth]{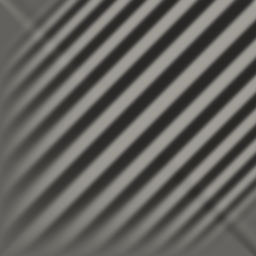}
  & \includegraphics[width=0.15\textwidth]{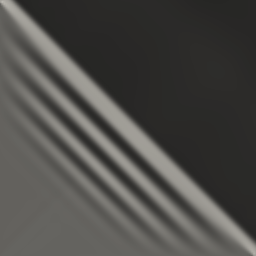}
  & \includegraphics[width=0.15\textwidth]{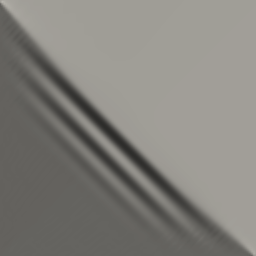}
  & \includegraphics[width=0.15\textwidth]{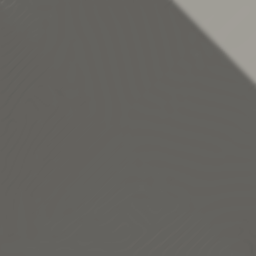}
  & \includegraphics[height=0.15\textwidth]{png/e2_legend.png}
\\
$c$
  & \includegraphics[width=0.15\textwidth]{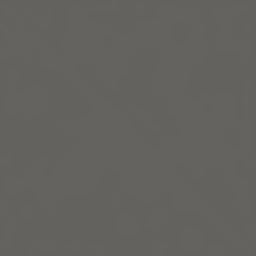}
  & \includegraphics[width=0.15\textwidth]{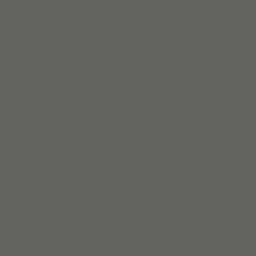}
  & \includegraphics[width=0.15\textwidth]{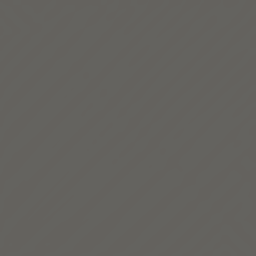}
  & \includegraphics[width=0.15\textwidth]{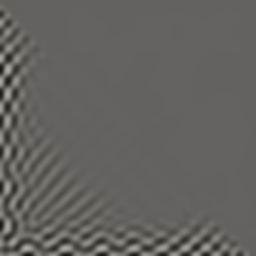}
  & \includegraphics[width=0.15\textwidth]{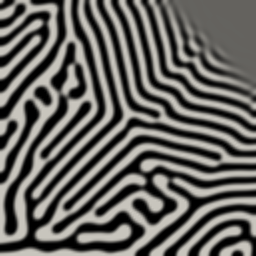}
  & \includegraphics[height=0.15\textwidth]{png/c_legend.png}
\\
  &   $t = 0.04$ & $t = 0.1$ & $t = 0.4$ & $t = 2$ & $t = 7$ &
\end{tabular}
\caption{The displacive order parameter $e_{2}$ and the diffusive order
parameter $c$ for the rapid cooling case ($\delta = 5 \times 10^{-2}$)
to $T_{\rm ext}= 0.45$.  At all snapshots, the temperature in the domain
is approximately~$0.45$.}
\label{fig:rapid_0.45}
\end{figure}

The final example is rapid cooling with $\delta = 5 \times 10^{-2}$ to
$T_{\rm ext} = 0.45$.  The time step was changed during the simulation,
and ranged between $\Delta t = 5 \times 10^{-4}$ and $\Delta t = 2
\times 10^{-1}$.  Snapshots of the order parameters $e_{2}$ and $c$
are presented in Figure~\ref{fig:rapid_0.45}.
As expected, martensite develops quickly. Interestingly, the domain
then undergoes a rearrangement between $t=0.1$ and $t=0.4$, in which
$e_{2}$ goes from an alternating pattern to a value close to zero in
the bottom left-hand region of the domain. It is likely that this is
due to the boundary constraint resisting deformation, and the driving
force for the square-to-rectangular transition being too small to
overcome this constraint.  In the top right-hand triangle, $e_{2}$ is
approximately constant at $t=0.4$.  At $T=0.45$, the driving force behind
the development of the alternating $e_{2}$ pattern is relatively weak,
whereas the energy associated with phase boundaries does not have a direct
temperature dependency.  The system undergoes another change between
$t=0.4$ and $t=2$, with the sign of $e_{2}$ in the upper right-hand
region changing.  On a longer time scale, diffusive processes develop,
first in the region where $e_{2} \approx 0$ and then progressing in into
the region where $e_{2}$ is not equal to zero, eventually replacing the
martensite phase. Noteworthy is the lack of structure in the diffusive
phase in the bottom left-hand triangular region, and the more coherent
lamella structure in the upper right-hand region.

As for the previous example, computing solutions to this problem robustly
and on a reasonable computational time scale is extremely challenging.

\section{Conclusions}
\label{sec:conclusions}

A phase field model that can simulate both diffusive and displacive
phase transitions has been presented. The model is developed in a
formal thermodynamic setting, with free-energy expressions associated
with distinct microstructural processes, the postulation of an energy
balance and demonstration that the classical entropy inequality is
satisfied. Various couplings terms introduce dependencies between
different processes.  The model includes both bulk energy and
phase boundary energy, which impacts on the fineness of the computed
microstructures.  Surface energies have been introduced via higher-order
spatial gradients of the relevant order parameters, and this demands
special care in the formulation of the thermodynamic balance laws. To
satisfy the classical point-wise form of the entropy inequality,
non-standard stress-like terms have been introduced to the energy
balance equation. These non-local terms are related to phase boundary
surface energy, and represent the work done by particles that neighbour
an arbitrary subdomain.

The differential equations that follow from the fundamental balance
laws involve the displacement field, the solute concentration and the
temperature field, with significant couplings between all equations. The
presence of surface energies in the model leads to a momentum balance
and mass balance (diffusion) equations that involve fourth-order
spatial derivatives. The presence of the higher-order derivatives is
problematic when developing a Galerkin finite element method for solving
the problem. To counter this, a sophisticated Galerkin method has been
formulated for the problem that imposes the classically required solution
regularity in a weak sense.  The computer model has been generated for a
large part automatically from a high-level specification, and is published
as supporting material.

It has been demonstrated through numerical simulations that the proposed
model can capture qualitatively a variety of observed phenomena,
including the formation of martensite twins, the development of
pearlitic structures and pearlite replacing martensite when a specimen
is held at a sufficiently high temperature for an extended period.
Phase transitions have been triggered via control of heat flux across
the boundary, but transformations can also be induced by mechanical
loading.  The application of the model with physically determined
parameters requires further investigation.  Such an investigation will
also necessitate further development of numerical solution strategies
to enable simulations on large domains and appropriate time scales to
be performed both accurately and within a tolerable simulation time.

\section*{Acknowledgements}

MM acknowledges the financial support of the Marco Polo Programme of the
University of Bologna and the hospitality of the Department of Engineering
at University of Cambridge.  The support of Prof.~Pier Gabriele Molari
throughout this work is gratefully acknowledged.  We also acknowledge
the work of Kristian B. {\O}lgaard on the code generation tools for
complicated equations, from which we have benefited.
\bibliography{references}
\bibliographystyle{elsarticle-harv}
\end{document}